\documentclass[twocolumn]{aastex63}

\usepackage[T1]{fontenc}
\usepackage{bm}        
\usepackage{amssymb, amsmath}   
\usepackage{cancel}
\usepackage{color}

\DeclareMathAlphabet{\mathsfit}{\encodingdefault}{\sfdefault}{m}{sl}
\SetMathAlphabet{\mathsfit}{bold}{\encodingdefault}{\sfdefault}{bx}{sl}
\newcommand{\vect}[1]{\bm{#1}}

\DeclareMathOperator{\sech}{sech}
\shorttitle{Observables from Magnetic Reconnection}
\shortauthors{Zhang et al.}

\begin{document}

\title{Radiation and Polarization Signatures from Magnetic Reconnection in Relativistic Jets--I. A Systematic Study}

\correspondingauthor{Haocheng Zhang}
\email{astrophyszhc@hotmail.com}

\author[0000-0001-9826-1759]{Haocheng Zhang}
\affiliation{Department of Physics and Astronomy \\
Purdue University \\
West Lafayette, IN 47907, USA}

\author{Xiaocan Li}
\affiliation{Dartmouth College\\
Hanover, NH 03750, USA}

\author{Dimitrios Giannios}
\affiliation{Department of Physics and Astronomy \\
Purdue University \\
West Lafayette, IN 47907, USA}

\author{Fan Guo}
\affiliation{Theoretical Division \\
Los Alamos National Lab \\
Los Alamos, NM 87545, USA}

\author{Yi-Hsin Liu}
\affiliation{Dartmouth College\\
Hanover, NH 03750, USA}

\author{Lingyi Dong}
\affiliation{Department of Physics and Astronomy \\
Purdue University \\
West Lafayette, IN 47907, USA}

\begin{abstract}
Blazars are relativistic magnetized plasma outflows from supermassive black holes that point very close to our line of sight. Their emission is nonthermal dominated and highly variable across the entire electromagnetic spectrum. Relativistic magnetic reconnection has been proposed as the driver of particle acceleration during blazar flares. While recent particle-in-cell (PIC) simulations have self-consistently studied the evolution of magnetic reconnection and particle acceleration therein, the resulting radiation signatures have not been systematically explored. In particular, the polarization signatures, which directly reflect the characteristic strongly dynamical magnetic field evolution during reconnection, have not been carefully investigated. In this paper, we present a systematic study of radiation and polarization signatures arising from magnetic reconnection in blazars, based on combined PIC and polarized radiation transfer simulations with various physical parameters. We identify a harder-when-brighter trend in the spectral evolution. Moreover, higher-frequency bands (ultraviolet to X-ray) tend to flare earlier than lower-frequency bands (infrared to optical) in the synchrotron spectral component. Most importantly, polarization signatures appear more variable with higher frequencies. We find that the temporal polarization variations strongly depends on the guide field strength. Specifically, reconnection with significant guide field component leads to very high polarization degree that contradict to typical blazar observations, while large polarization angle rotations are unique signatures of magnetic reconnection between nearly anti-parallel magnetic field lines. These rotations are at least $90^{\circ}$ and can extend to $>180^{\circ}$, and they may rotate in both directions. These results imply that blazars that have shown large polarization angle rotations intrinsically have more nearly anti-parallel magnetic field morphology.
\end{abstract}

\keywords{galaxies: jets --- radiation mechanisms: non-thermal --- magnetic reconnection --- polarization}


\section{Introduction} \label{sec:intro}

Relativistic jets from active galactic nuclei (AGN) are among the most extreme astrophysical phenomena in the universe. These plasma jets are powered by the accretion of the supermassive black hole at the center of AGN. Blazars are a kind of AGN whose jet directs very close to our line of sight. Their emission is characterized by a two-hump shaped spectral energy distribution (SED) and dominated by nonthermal radiation processes \citep[for a recent review on blazars, see e.g.,][]{Boettcher19}. The low-energy hump is dominated by synchrotron emission of ultra-relativistic electrons. This is evident by the observed polarization degree (PD), which is consistent with synchrotron emission in a partially ordered magnetic field \citep{Pushkarev05,Zhang15}. The origin of the high-energy hump can be either leptonic or hadronic. In the former case, the high-energy emission comes from the Compton scattering of low-energy photons by the same electrons that produce the low-energy synchrotron component \citep{Marscher85,Maraschi92,Dermer92,Sikora94}. In the latter case, the X-ray to $\gamma$-ray emission is due to proton synchrotron and/or hadronic cascades \citep{Mannheim93,Aharonian00,Mucke03}. The recent very high energy neutrino detection that is simultaneous with a blazar flare provides the first evidence that the high-energy hump may be of hadronic origins \citep{IceCube18}. It has been suggested that high-energy polarimetry, especially the MeV $\gamma$-ray polarization signatures, can diagnose whether the high-energy spectral component is of leptonic or hadronic origins \citep{Zhang13,Paliya18,Zhang19,Rani19,McEnery19}.

The entire blazar spectrum can be highly variable, with the $\gamma$-ray bands can flare in as short as a couple of minutes \citep{Aharonian07,Albert07,Ackermann16}. Such extreme flares require very fast and efficient particle acceleration within a very small region in space. This localized region is often referred to as the blazar zone. Blazar flares are often interpreted as results of internal shocks that accelerate a large amount of nonthermal particles via the diffusive shock acceleration \citep[e.g.,][]{Marscher85,Boettcher10,Boettcher19b}. However, shock models have some troubles in explaining, for instance, the very fast variability in $\gamma$-rays. In recent years, there is an increasing interest on the magnetic-driven jets, where blazar flares are considered to be driven by magnetic reconnection in the emission region \citep{Giannios13,Sironi15,Zhang18,Giannios19}.

Magnetic reconnection is a plasma physical process where oppositely directed magnetic field lines rearrange their topology and release a large portion of their magnetic energy. It can be an efficient way to accelerate nonthermal particles if the blazar emission region is considerably magnetized. Numerical simulations including particle-in-cell (PIC) simulations have shown that magnetic reconnection can accelerate both electrons and protons into power-law spectra \citep{Sironi14,Guo14,Guo16,Guo19,Werner16,Werner18,li18,li19,kilian20}. Depending on how much the reconnection region is magnetized, the nonthermal particle spectra can be either hard or soft, which are consistent with the observed blazar radiation spectra \citep{Guo15,Sironi15,Petropoulou19}. Recently, several works have studied the light curves arising from magnetic reconnection, which appear overall consistent with blazar observations \citep{Deng16,Petropoulou16,Christie19}. However, those works have not yet systematically studied the observable behaviors and trends arising from magnetic reconnection. The issue lies in mainly three aspects. First, magnetic reconnection is a highly dynamical plasma process, but its time-dependent radiation signatures have not been well studied. Second, the guide field strength that is crucial to the reconnection dynamics can also strongly affect radiation signatures, but previous works have not studied its effects on radiation. Third, polarization signatures, which can directly reflect the intrinsic magnetic field evolution in the reconnection region, remain largely unexplored.

Polarimetry can probe the magnetic field morphology and evolution in astrophysical systems. Since the low-energy component of the blazar spectrum is dominated by synchrotron emission, optical polarization signatures can directly reflect the magnetic field evolution in the blazar flaring regions \citep[see][for a recent review]{Zhang2019b}. Although the optical polarization often fluctuates erratically at a low PD during quiescent states, it can reach higher PD and become strongly variable when the blazar is flaring \citep{Smith09,Dammando11,Ikejiri11}. Particularly, observations have detected large optical polarization angle (PA) swings simultaneously with multi-wavelength blazar flares, indicating significant magnetic field evolution \citep{Marscher08,Larionov13,Blinov15}. Very interestingly, the RoboPol project has noticed several statistical trends in the optical polarization signatures, such as the PA swings are mostly correlated to $\gamma$-ray flares \citep{Angelakis16,Blinov18}. These systematic trends indicate that the polarization variations in blazars originate from physical processes that are not described by completely stochastic random walks \citep{Kiehlmann17}. It has also been reported in several sources that PA swings can reach far beyond $\sim 180^{\circ}$, and can take place in both directions in the same blazar \citep[e.g.,][]{Marscher10,Morozova14,Chandra15}. Furthermore, the PD generally drops during the PA swings \citep{Blinov16}. These behaviors indicate that the physical driver of blazar flares with PA swings can strongly alter the magnetic field morphology in the blazar emission region.

PIC simulations have shown that the reconnection layer can fragment forming a large number of moving plasmoids of different sizes. Such magnetic field morphology and evolution are characteristic to magnetic reconnection and unlike to those expected in shocks and turbulence. As a first attempt to identify characteristic signatures of magnetic reconnection, \cite{Zhang18} have shown that magnetic reconnection in the blazar environment can produce large optical PA swings (beyond $180^{\circ}$), and the PA can swing in both directions during flares. The paper suggests that the PA swings intrinsically originate from the secondary reconnection due to plasmoid mergers, which are unique to magnetic reconnection events. This illustrates a promising observable constraint to identify and diagnose magnetic reconnection events in blazars.

Inspired by the study of \cite{Zhang18}, we perform a series of combined  PIC simulations with polarization-dependent radiation transfer simulations. We aim to systematically study the effects on radiation for several key physical parameters, namely, the guide field strength, magnetization factor, radiative cooling, upstream temperature, and observational frequencies. Additionally, we perform our simulation in a larger simulation box to examine if the radiation signatures are consistent with larger physical sizes. Our goal is to identify general observable patterns of multi-wavelength light curves and optical polarization signatures arising from magnetic reconnection in the blazar emission region. In addition, we want to diagnose if there exist specific signatures or trends that various physical parameters of the reconnection layer can affect the radiation and polarization signatures. Section \ref{sec:setup} describes our numerical setup, section \ref{sec:results} presents the radiation and polarization signatures arising from reconnection with guide fields, section \ref{sec:parameterstudy} performs additional parameter studies, and section \ref{sec:observation} discusses the observational implications. We conclude our paper in section \ref{sec:discussion}.

\section{Simulation Setup \label{sec:setup}}

The goal of this paper is to systematically study radiation and polarization signatures resulting from magnetic reconnection events in relativistic jets. We assume that the reconnection happens in the blazar zone environment. We start our simulation from a pre-existing current sheet with various initial physical conditions. Then we will investigate how radiation and polarization signatures may depend on the initial physical parameters. The general setups of combined PIC and polarized radiation transfer simulations are the same as \cite{Zhang18}. In the following, we briefly summarize the setups and describe any additional components to those in \cite{Zhang18}.

\subsection{PIC Setup}

We perform 2D PIC simulations in the $x$--$z$ plane using the \texttt{VPIC} code \citep{Bowers08}, which solves the Maxwell's equations and the relativistic Vlasov equation. The simulations start from a magnetically-dominated force-free current sheet, $\vect{B}=B_0\tanh(z/\lambda)\hat{x}+B_0\sqrt{\sech^2(z/\lambda)+B_g^2/B_0^2}\hat{y}$, where $B_g$ is the strength of the guide field (magnetic field component perpendicular to the reconnecting magnetic field). We set the half-thickness $\lambda$ of the current sheet to be $0.6\sqrt{\sigma_e}d_{e0}$ in order to have enough particles in the current sheet for carrying electric current to satisfy the Ampere's law, where $d_{e0}=c/\omega_{pe0}$ is the nonrelativistic electron inertial length, $\omega_{pe0}=\sqrt{4\pi n_ee^2/m_e}$ is the nonrelativistic electron plasma frequency, and $\sigma_e=B_0^2/(4\pi n_em_ec^2)$ is the cold electron magnetization parameter \citep{Sironi14,Guo14}. The initial particle distributions are Maxwell–J\"uttner distributions with uniform density $n_0$ and temperature $T_e=T_i$. The simulation assumes an electron-ion plasma with realistic mass ratio $m_i/m_e=1836$. We expect that our radiation and polarization results should hold for pair plasma and electron-positron-proton plasma, because the leptons consume a significant portion of the dissipated magnetic energy, and the reconnection dynamics and evolution are generally similar \citep{Petropoulou19}.  We use $100$ electron-ion pairs in each cell. The simulation box size is $2L\times L$ in the $x$--$z$ plane. We employ periodic boundary conditions in the $x$-axis for both fields and particles, while in the $z$-axis the boundaries are conductive for fields but reflect particles. We insert a long-wavelength perturbation to trigger the magnetic reconnection, which creates a dominating reconnection point located at the center of the simulation box \citep{Birn01}. The radiative cooling effects are important for blazars. Here we mimic the cooling effect by implementing a radiation reaction force $\vect{g}$, which can be simplified as a continuous friction force for ultra-relativistic particles~\citep{Cerutti12, Cerutti13},
\begin{align*}
  \vect{g} & = -\frac{\mathcal{P}_\text{rad}}{c^2}\vect{v} \\
  & = -\frac{2}{3}r_e^2\gamma\left[\left(\vect{E}+
  \frac{\vect{u}\times\vect{B}}{\gamma}\right)^2 -
  \left(\frac{\vect{u}\cdot\vect{E}}{\gamma}\right)^2\right]\vect{u}-\frac{4}{3}\sigma_T\gamma \mathcal{U}_{\star}\vect{u},
\end{align*}
where $\vect{u}=\gamma \vect{v}/c$ is the four-velocity, $\mathcal{P}_\text{rad}$ is the radiation power, $r_e=e^2/m_ec^2$ is the classical radius of the electron, and $\mathcal{U}_{\star}$ is the photon energy density \citep[also see,][for implementation details in \texttt{VPIC}]{Zhang18}. Since nonthermal electrons in the blazar emission region are highly relativistic, the nonrelativistic terms in the radiation reaction force is not included in our simulations \citep{Cerutti2017}.

We choose simulation parameters according to the physical conditions inferred by spectral modeling of the observations of flat-spectrum radio quasars (FSRQs), which exhibit the strongest flux and polarization variability~\citep{Ackermann16,Angelakis16}. We choose the plasma thermal temperature $T_e=100m_ec^2$ for our default run. This is below the low-energy cutoff \citep[ranging from hundred to thousand $m_ec^2$, see e.g.,][]{Boettcher13,Paliya18} of the nonthermal electron spectra inferred for FSRQs, enabling the simulations to capture the formation of nonthermal electron spectra. Spectral fitting for FSRQs suggests that the high-energy cutoff of the electron spectrum is $\gamma\sim 10^4$ \citep{Boettcher13,Paliya18}. Since the high-energy cutoff is roughly equal to the electron magnetization factor $\sigma_e$~\citep[e.g.][]{Guo14}, we choose $\sigma_e=4\times 10^4$ for our default run. This corresponds to a total magnetization $\sigma_0\approx(m_e/m_i)\sigma_e\approx 22$.

The radiative cooling plays an important role in blazar radiation and polarization signatures. However, since the physical scale of the PIC simulation is very small compared to the realistic blazar zone, we need to normalize the cooling rate by the acceleration rate in PIC simulations so that they can produce similar light curves as observations. Although the exact particle acceleration mechanism in magnetic reconnection is a very complicated issue and beyond the scope of this paper, we only need the acceleration rate as a normalization for the cooling rate. For simplicity, we take the results from recent analyses to PIC simulations on the acceleration rate \citep{Guo14,Guo15,Li2019}, where have shown that the acceleration rate $\alpha=\dot{\gamma}/\gamma$ scales with $\sqrt{\sigma_e}$. Based on their measurement, we determine that the particle acceleration rate is approximately $\alpha=\dot{\gamma}/\gamma\approx\sqrt{\sigma_e}\omega_{pe0}/2000$. The particle cooling time scale is given by $\tau_\text{cool} = 3t_0/(2\gamma\sigma_e\tilde{r}_e)$~\citep{Zhang18}, where $t_0=\omega^{-1}_{pe0}$ and $\tilde{r}_e=r_e/ct_0$. For FSRQs, the cooling time $\tau_\text{cool}$ is usually longer than the acceleration time. Previous works have used the $\gamma_{rad}$ where the acceleration and cooling time scales are equal to normalize the cooling effects \citep{Cerutti2016}. However, this choice often leads to very strong cooling, which does not fit with typical FSRQ parameters \citep{Yuan2016}. To capture the cooling effects in simulation time, we adjust $\tau_\text{cool}$ so that $C_{10^4}\equiv\alpha\tau_\text{cool}=200$ for electrons with $\gamma=10^4$ in the default run. As shown in the following, the resulting radiation and polarization signatures are in good agreement with observations. In principle, the cooling due to synchrotron and Compton scattering are different. However, for typical FSRQs, the two cooling time scales have very similar expressions, except that the former is proportional to the magnetic energy density, while the latter is proportional to the photon energy density. The exact ratio of the two energy densities only matters when we study the multi-wavelength radiation signatures. In this paper, we focus on the synchrotron signatures, thus we only consider the synchrotron cooling term in our simulations.

For the default run, $L_x=2L=1.6\times 10^4d_{e0}$ and $L_z=L=8\times 10^3d_{e0}$, which is normalized to $\sim 8.5\times 10^{10}~\rm{cm}$ in typical FSRQs ($B\sim 0.1~\rm{G}$, $n_e\sim 0.01~\rm{cm^{-3}}$). While this is much smaller than the typical blazar emission region ($\sim 10^{16}$ cm), we find that the general plasmoid dynamics are qualitatively the same with domain size $2\times$ larger than the present case (refer to the BS1 simulation in Section \ref{sec:parameterstudy}). Since the key mechanism in producing radiation signatures is the plasmoid coalescence/merger, as demonstrated in this paper, this suggests that the underlying process is robust even on the macroscopic scales over which blazar flares take place. We choose a simulation grid size of $4096\times2048$ for the default run, so that the cell sizes $\Delta x=\Delta z\sim 0.32d_e$ can resolve the thermal electron inertial length $d_e=\sqrt{\gamma_0}d_{e0}$, where $\gamma_0=1+3T_e/2m_ec^2\sim 150$. We use the same $\Delta x$ and $\Delta z$ for all runs. Table~\ref{tbl:list_runs} shows the simulation parameters for all the 11 runs. We name the default run DEF, which has a guide field $B_g/B_0=0.2$. We vary the guide field from 0.0 to 1.0 (GF1--4) to study how the guide field changes electron acceleration and radiation signatures. We compare the default run with runs with different magnetization factors $\sigma_e$ (MF1 and MF2) and cooling factors $C_{10^4}$ (CF1 and CF2), higher upstream plasma temperature (UT1), and larger box size (BS1).


\begin{deluxetable}{cccccc}
  \tabletypesize{\scriptsize}
  \tablecaption{List of PIC simulations\label{tbl:list_runs}}
  \tablewidth{0pt}
  \tablehead{
  \colhead{Run} &
  \colhead{$B_g/B_0$} &
  \colhead{$\sigma_e$} &
  \colhead{$T_e/m_ec^2$} &
  \colhead{$C_{10^4}$} &
  \colhead{$L_x/d_{e0}$}
  }
  \startdata
 DEF & 0.2 & $4.0\times10^4$ & 100 & 200 & $1.6\times10^4$ \\
 GF1 & 0.0 & $4.0\times10^4$ & 100 & 200 & $1.6\times10^4$ \\
 GF2 & 0.4 & $4.0\times10^4$ & 100 & 200 & $1.6\times10^4$ \\
 GF3 & 0.6 & $4.0\times10^4$ & 100 & 200 & $1.6\times10^4$ \\
 GF4 & 1.0 & $4.0\times10^4$ & 100 & 200 & $1.6\times10^4$ \\
 MF1 & 0.2 & $10^4$          & 100 & 200 & $1.6\times10^4$ \\
 MF2 & 0.2 & $1.6\times10^5$ & 100 & 200 & $1.6\times10^4$ \\
 CF1 & 0.2 & $4.0\times10^4$ & 100 & 100 & $1.6\times10^4$ \\
 CF2 & 0.2 & $4.0\times10^4$ & 100 & 400 & $1.6\times10^4$ \\
 UT1 & 0.2 & $4.0\times10^4$ & 400 & 200 & $1.6\times10^4$ \\
 BS1 & 0.2 & $4.0\times10^4$ & 100 & 200 & $3.2\times10^4$
  \enddata
  \tablecomments{All PIC simulation parameters. There are in total 11 PIC runs: DEF is the default parameter set; GF1-4 are for different guide fields; MF1-2 are for different magnetization factors; CF1-2 are for different cooling factors; UT1 is for a higher upstream temperature; and BS1 is for a larger box size. We consider the following parameters: $B_g/B_0$ is the ratio between the guide field component and the anti-parallel magnetic field component; $\sigma_e$ is the electron magnetization factor; $T_e$ is the upstream temperature of particles; $C_{10^4}\equiv\alpha\tau_\text{cool}$ for electrons with $\gamma=10^4$; and $L_x/d_{e0}$ is the simulation box width in the unit of $d_{e0}$.}
\end{deluxetable}

\subsection{Radiation Transfer Setup}

Since the reconnection simulation is performed in the $x$--$z$ plane in the \texttt{VPIC} code, we fix our line of sight in the comoving frame of the simulation box along the $y$-axis. This is because in our 2D PIC simulations, the evolution and morphology of the guide field component, which is the magnetic field component in the $y$ direction, is not resolved. Since the synchrotron emission only depends on the magnetic field perpendicular to the line of sight, setting the line of sight along $y$-axis can eliminate the effects of untracked guide field distribution and evolution on the synchrotron radiation signatures. We choose that the reconnection layer is moving in the $z$-direction with a bulk Lorentz factor $\Gamma=10$ in the observer's frame for all our simulations. Therefore, the Doppler factor in the observer's frame is $\delta=\Gamma=10$, which is a typical number for blazars. We normalize the initial anti-parallel magnetic field components in the reconnection plane (i.e., without the guide field) to be $0.1~\rm{G}$, which is a typical value found in the leptonic blazar spectral fitting \citep{Boettcher13,Paliya18}.

To obtain the particle distributions and magnetic fields, we reduce every $16\times 16$ PIC cells into one radiation transfer cell. We find this resolution is adequate to capture all relevant radiation features and provide enough statistics to obtain smooth particle spectra in each radiation transfer cell. We divide the particle kinetic energy $(\gamma-1)m_ec^2$ evenly into 100 bins in logarithmic space between $10^{-4}$ to $10^6$. Then we obtain the particle spectra by counting the number of particles in each energy bin. We calculate the magnetic field in the radiation transfer cell by averaging those in the $16\times16$ PIC cells. As we can see in all snapshots in Section \ref{sec:results} and \ref{sec:parameterstudy}, the magnetic field does not show very sharp changes on very small scales. Additionally, plasmoids smaller than $16\times 16$ PIC cells have very limited number of nonthermal particles within. Thus averaging the magnetic field in the $16\times16$ PIC cells does not lose any major observable signatures. Since we use a periodic boundary in the PIC simulations, we find that the motion of plasmoids in our simulations is generally non-relativistic. Therefore, we do not include any local Lorentz factor in our radiation transfer simulations.

We use the \texttt{3DPol} code developed by \cite{Zhang14} to perform radiation transfer simulations. This code is a polarization-dependent radiation transfer code for synchrotron emission. It evaluates the Stokes parameters of the synchrotron emission (Stokes parameters represent the polarization status in the emission) from each cell in the simulation, so as to include all linear polarization signatures, based on the magnetic field and particle distributions, which are obtained from the PIC simulations. It then traces the emission beams to the plane of sky, and add up all emission in the same cell on the plane of sky within the same frequency band that arrive at the same time step. Since the line of sight in our simulation is set to be along $y$-axis, the plane of sky is then parallel to the $x$--$z$ plane. The code has time-, space-, and frequency-dependencies. A key feature of the \texttt{3DPol} code is that it allows us to get the polarized emission maps at every time step. This feature can illustrate the surface brightness and polarization distributions in the simulation domain, so that we can pinpoint the plasma dynamics with resulting radiation behaviors.

\section{Radiation and Polarization Signatures from Magnetic Reconnection \label{sec:results}}

\cite{Zhang18} have presented the radiation and polarization signatures arising from magnetic reconnection between perfectly anti-parallel magnetic fields, in which we have found strong PA swings. Here we present additional simulations to understand how different physical parameters may affect the radiation and polarization signatures. We recognize that in relativistic jets, current sheets may form via magnetic instabilities/turbulence or striped jet \citep{Begelman98,Giannios06,Giannios19}. In the former case, we expect that the reconnecting magnetic field lines are unlikely perfectly anti-parallel, but have a finite guide field component; in the latter case, the oppositely oriented magnetic stripes are initially formed at the central engine due to magnetic irregularities that are advected into the jet. Depending on the magnetic structure at the central engine, the reconnecting magnetic field lines may or may not be perfectly anti-parallel. Guide fields can considerably affect the magnetic reconnection dynamics \citep{Lyubarsky05,Ball19,Rowan19,Liu15,Liu19}. Furthermore, different observational frequencies may also lead to different spectral and temporal patterns in both radiation and polarization signatures. In this section, we present general radiation and polarization signatures from magnetic reconnection with a finite guide field, and study the effects of observational frequency and guide field strength.

\subsection{General Temporal and Spectral Behaviors}

\begin{figure}
    \centering
    \includegraphics[width=\linewidth]{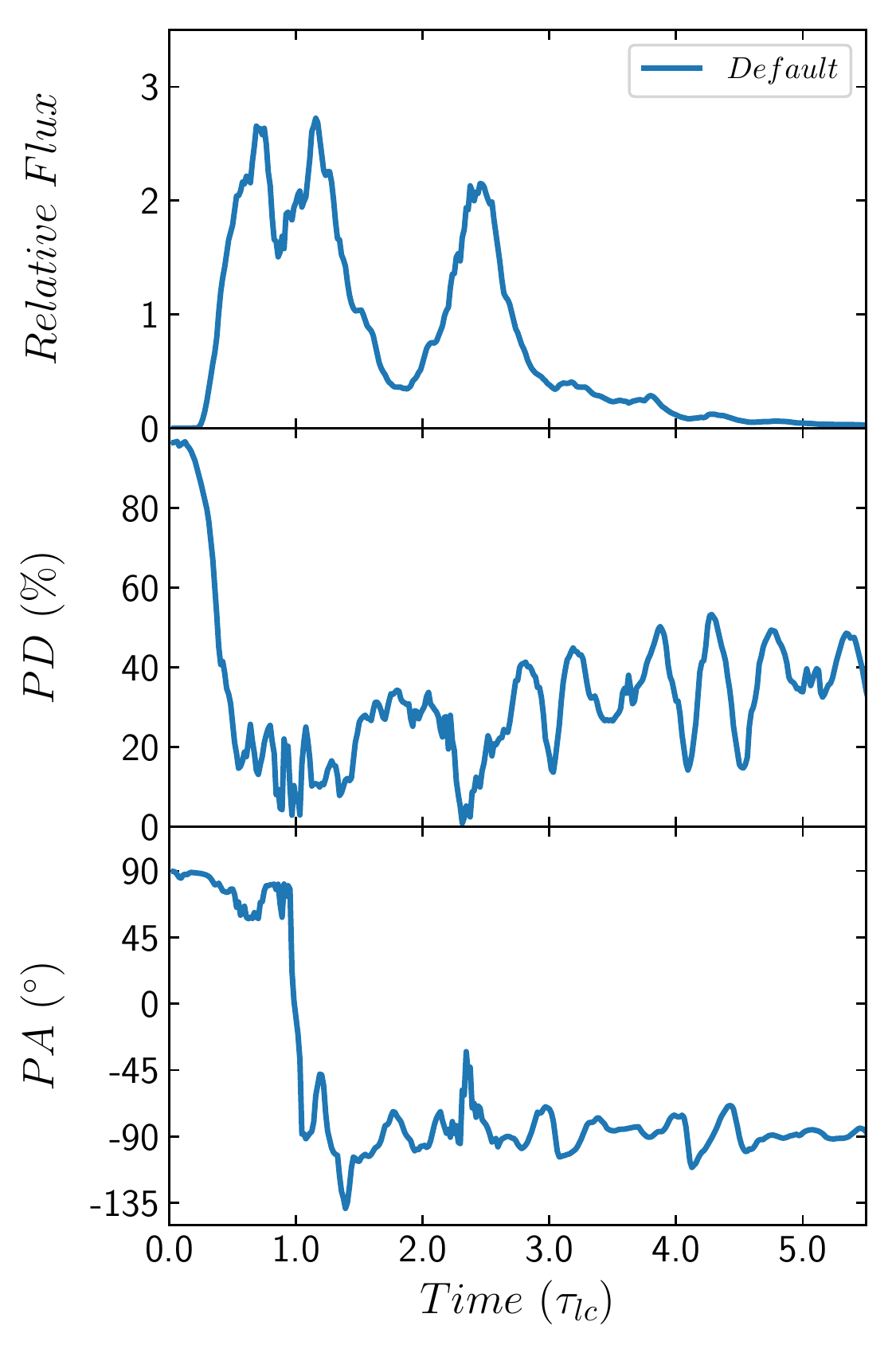}
    \caption{From top to bottom panels are optical light curve, PD, and PA for our default setup. Time is in the unit of light crossing time ($\tau_{lc}\equiv L_x/c$) of the simulation box length. The light curve is plotted in relative flux. All results are presented in the observer's frame.}
    \label{fig:default}
\end{figure}

\begin{figure}
    \centering
    \includegraphics[width=\linewidth]{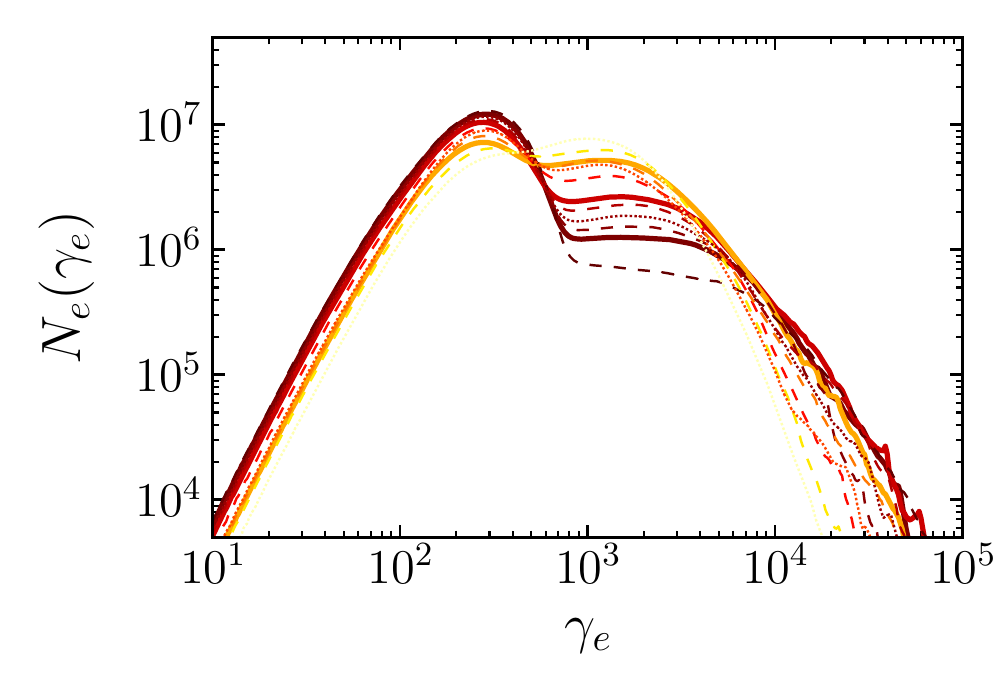}
    \includegraphics[width=\linewidth]{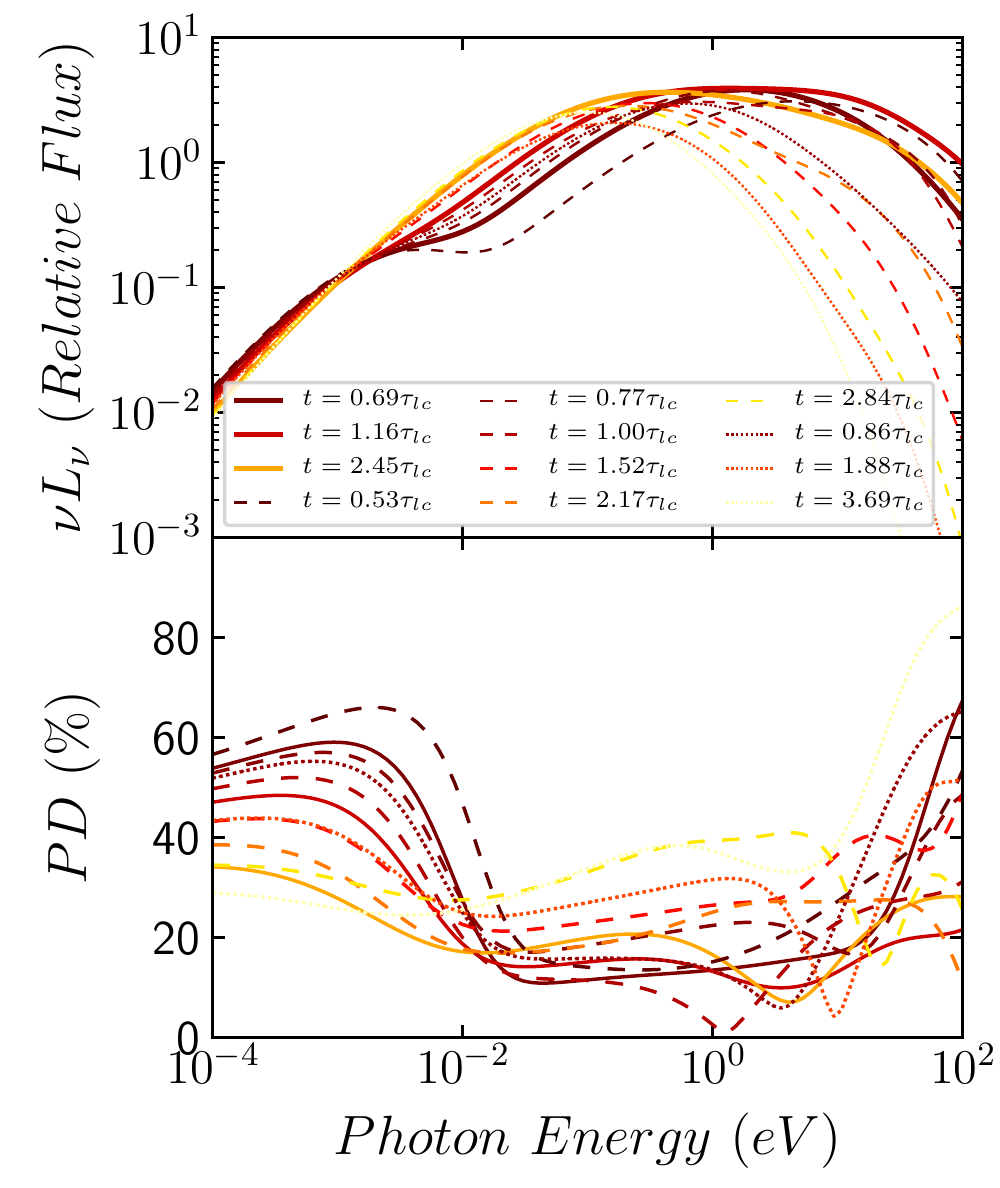}
    \caption{From top to bottom panels are snapshots of particle spectra, SEDs of the synchrotron component, and frequency-dependent PDs. SEDs are plotted in relative flux. The first three bold solid lines are snapshots chosen at the peak of the light curve, the following six dashed lines are in between, the last three dotted lines are snapshots at the valleys (the readers can refer to Figure \ref{fig:default} for exact location). All results are presented in the observer's frame.}
    \label{fig:defaultspec}
\end{figure}

We set up the reconnection layer with an initial electron magnetization factor of $\sigma_e=4\times10^4$ with a guide field of $B_g/B_0=0.2$ (see Table \ref{tbl:list_runs} DEF for physical parameters). Since we use the real mass ratio between protons and electrons, this yields a total magnetization factor of $\sigma_0\sim 22$. Figures \ref{fig:default} and \ref{fig:defaultspec} present the temporal and spectral radiation signatures, respectively.

We trigger the magnetic reconnection with a small initial perturbation. From the light curve (Figure \ref{fig:default} top panel), we can see that the reconnection starts to accelerate particles to high Lorentz factors around $t\sim 0.3\tau_{lc}$, where $\tau_{lc}\equiv L_x/c$ is the light crossing time. Figure \ref{fig:default3070160256} plots snapshots of spatial distributions of magnetic field strength, nonthermal particles, and polarized emission maps in the simulation domain. The four snapshots correspond to the rising phase of the first flare in the light curve, the peak of the second and third flare, and the saturation of magnetic reconnection. We can clearly observe that soon after the trigger of reconnection, the reconnection layer fragments into a series of plasmoids (Figure \ref{fig:default3070160256} left three panels). These plasmoids are quasi-circular structures in our 2D simulation. They are pervaded by magnetic field loops with high density of nonthermal particles. The direction of plasmoid magnetic field loops is clockwise, due to the initial choice of the magnetic field topology, where the upper half of the simulation domain has magnetic component in the reconnection plane along $+x$ direction, while the lower half has that along the $-x$ direction. As we can see in the following, this choice of does not affect the reconnection development or the radiation and polarization signatures.

\subsubsection{Temporal Patterns}

We find that the plasmoids produced in the reconnection layer generally move away from the main X-point. But they can have different bulk speeds, so that they may collide and merge into each other. Since all plasmoids produced from the primary reconnection have clockwise magnetic fields (if initial morphology were reversed, then magnetic fields in the plasmoids would all point to the counterclockwise direction), when they collide and merge, they form a current sheet at the contact region, and trigger secondary reconnection (see Figure \ref{fig:default54586266} for the evolution of a merger event). The first flare is because the primary reconnection accelerates a large amount of nonthermal particles. We find that the small fluctuations/spikes on the light curves originate from mergers of smaller plasmoids. On the other hand, the second and third flares are due to large plasmoid mergers. After the third flare, a large amount of the magnetic energy in the reconnection layer has depleted, and there is only one remaining large plasmoid, due to our periodic boundary condition, that passively cools. Therefore, there are no additional flares afterwards.

\begin{figure*}
    \centering
    \includegraphics[width=\textwidth]{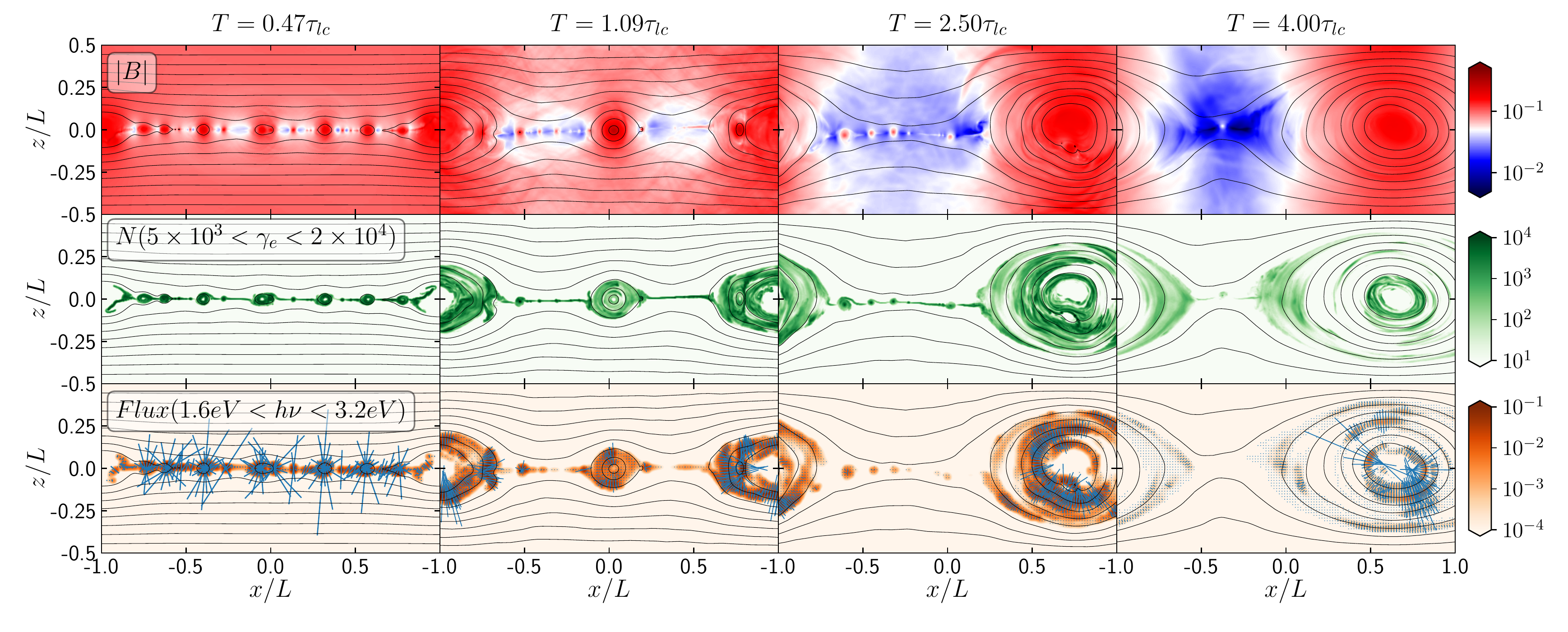}
    \caption{Snapshots of the reconnection evolution for the default setup. The first row is the magnetic field strength, which is plotted using the original PIC simulation resolution, $4096\times 2048$. The second row is the spatial distribution of nonthermal particles within the given Lorentz factor range. It is plotted after the reduction of $16\times 16$ PIC cells, at the resolution of $256\times 128$. The third row is the resulting polarized emission maps in the given observational band. The black lines in all panels trace the magnetic field lines. The blue dashes in the third row represent the local polarized flux. Their lengths are proportional to the polarized flux amplitude, and their directions illustrate the local PA.}
    \label{fig:default3070160256}
\end{figure*}

\begin{figure*}
    \centering
    \includegraphics[width=\textwidth]{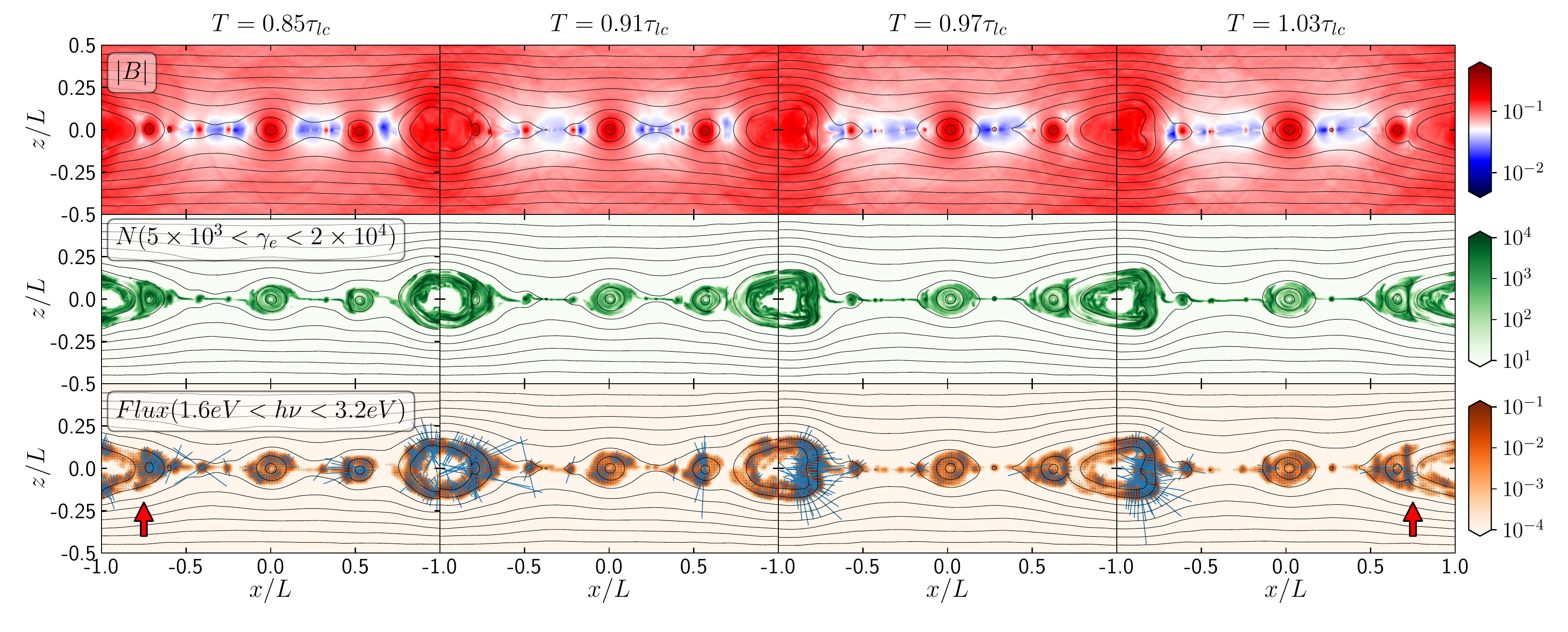}
    \caption{Same as Figure \ref{fig:default3070160256}, but these snapshots trace the evolution of a large plasmoid merger event (on the left end of the simulation domain marked by red arrow) in the default run. In the last snapshot, there is a large plasmoid merger starting on the right end of the simulation domain (also marked by the red arrow).}
    \label{fig:default54586266}
\end{figure*}

Before the reconnection starts, since the PA has $180^{\circ}$ ambiguity, the initial anti-parallel magnetic field components appear the same as a uniform field in polarization signatures. For the same reason, the clockwise and counterclockwise magnetic field loops in the plasmoids look the same as rings of magnetic field lines. Additionally, only the magnetic field components that are perpendicular to the line of sight contribute to synchrotron emission. Therefore, we see nearly $100\%$ PD at the beginning of the simulation. Soon after the reconnection starts, the reconnection layer fragments into many plasmoids, making the overall magnetic field morphology very disordered. Thus we observe that the PD quickly drops to $\lesssim 20\%$ when the flux becomes considerable (Figure \ref{fig:default} top and middle panels). Figure \ref{fig:default3070160256} shows that significant amount of polarized emission comes from plasmoids, which have high density of energetic particles and stronger magnetic fields. We can see that the polarize emission on the plasmoids has the PA generally perpendicular to the magnetic field loops. Given that the magnetic field of plasmoids are of circular shape and they frequently collide into each other, the overall magnetic field morphology during reconnection is disordered and variable. Since the reconnecting magnetic field lines are in the $\pm x$ directions, the overall emission has an excess contribution from the $\pm x$ magnetic field components. As a result, we observe that PD remains at $\lesssim 20\%$ and PA fluctuates around a mean value at $\pm 90^{\circ}$ (they are the same angle due to the $180^{\circ}$ ambiguity, representing $\pm x$ magnetic field components).

The situation changes when large plasmoids merge into each other. 
Since plasma is strongly compressed when plasmoids merge~\citep{Li2018Roles}, the secondary reconnection at the merging site accelerates more nonthermal particles. This makes its emission to dominate over that from other parts of the reconnection layer. Consequently, the temporal PD and PA variations during this period represent the evolution of the plasmoid merger. Figure \ref{fig:default54586266} tracks a plasmoid merger event on the left end of the simulation box during the large PA rotation between $t=0.85 \tau_{lc}$ and $t=1.05\tau_{lc}$ (the rising phase of the second flare). Clearly, we observe a large amount of nonthermal particles accelerated at the contact region. These newly accelerated particles can stream along the magnetic field lines of the two merging plasmoids, thus light up the magnetic field morphology at their location. However, particles that are streaming clockwise and counterclockwise are not of the same amount. This is very similar to the primary reconnection layer. As we can see in Figure \ref{fig:default3070160256} and snapshots of all other simulations, plasmoids and nonthermal particles that are produced to the left are generally not symmetric to those at the right. Consequently, at the secondary reconnection layer in the larger plasmoid mergers, if there is a considerable difference in the flow of particles between the clockwise and counterclockwise directions, the PA will show a smooth rotation representing the dominating direction. On the other hand, if the two outflows are comparable in the secondary magnetic reconnection, we do not expect a PA rotation. Nevertheless, in either situation there should be strong particle acceleration at major plasmoid mergers, so that we always expect a flare event. As we can see in the first three snapshots in Figure \ref{fig:default54586266}, the large plasmoid merger on the left side of the reconnection layer has more nonthermal particles in the counterclockwise direction, thus the PA makes a smooth and fast swing from $90^{\circ}$ to $-90^{\circ}$. At the last snapshot, there is another large plasmoid merger starting on the right. Although this one also has a counterclockwise preference, its starting PA position is different from the ending PA position of the previous one. This leads to a jump of PA at $t=1.05\tau_{lc}$, followed by a continuous PA swing to $-135^{\circ}$. On the other hand, the large plasmoid merger that leads to the third flare gives similar amount of nonthermal particles streaming in both directions. Therefore, we do not observe a PA swing associated with this flare.

\subsubsection{Spectral Properties}

Magnetic reconnection quickly accelerates electrons into a power-law distribution. Due to the synchrotron cooling, the particle spectra exhibit a broken power-law shape (Figure \ref{fig:defaultspec} top panel). The optical emission plotted in Figure \ref{fig:default} is beyond the cooling break. Very interestingly, we observe an overall harder-when-brighter trend for the cooling spectra (see Figure \ref{fig:defaultspec} middle panel, solid lines are always harder than dotted lines, while dashed lines are in between). This is because at the flare peak, where the primary reconnection and later on the large plasmoid mergers have the highest efficiency, the accelerated particle power-law distribution is very hard. It is also evident from Figure \ref{fig:defaultspec} that the part of the nonthermal particle spectra that do not suffer from strong cooling ($10^3<\gamma_e<4\times 10^3$) are very hard. Then the radiative cooling gradually softens the spectra after the flare peak, which results in an overall harder-when-brighter trend. We also find that the PD tends to be lower at the flare peak than the lower flux states (Figure \ref{fig:defaultspec} bottom panel). This is because there are more plasmoid production and mergers when the flux is higher, leading to more disordered magnetic field morphology in the reconnection layer. We observe that the PD is very high towards low and high ends of the spectra. The high energy end is easily understandable, as there are very few very high energy particles. The high PD at lower energies, however, is due to the fact that they originate from upstream thermal particles. These particles occupy the entire reconnection layer, including the very ordered magnetic field structure at the top and bottom of the reconnection layer. Nonetheless, this part of the emission is not expected to be observed, because they show very low flux and never flare during the reconnection development.

\subsection{Effects of Observational Frequencies}

\begin{figure}
    \centering
    \includegraphics[width=\linewidth]{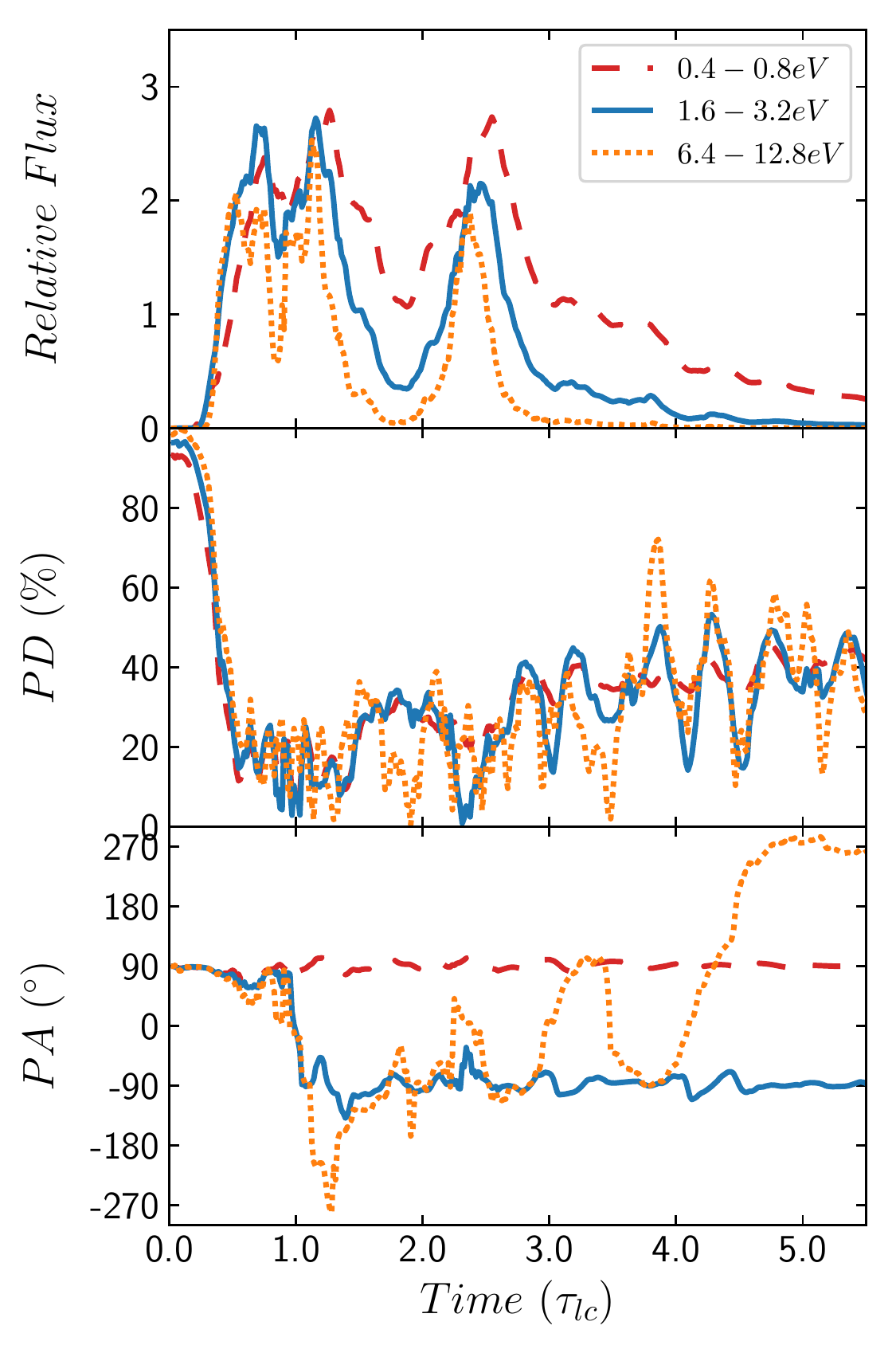}
    \caption{From top to bottom are the light curves, temporal PD and PA for three different bands in infrared, optical, and ultraviolet for the default run, which are four times apart in photon energies as shown in the legend.}
    \label{fig:obsband}
\end{figure}

\begin{figure*}
    \centering
    \includegraphics[width=\textwidth]{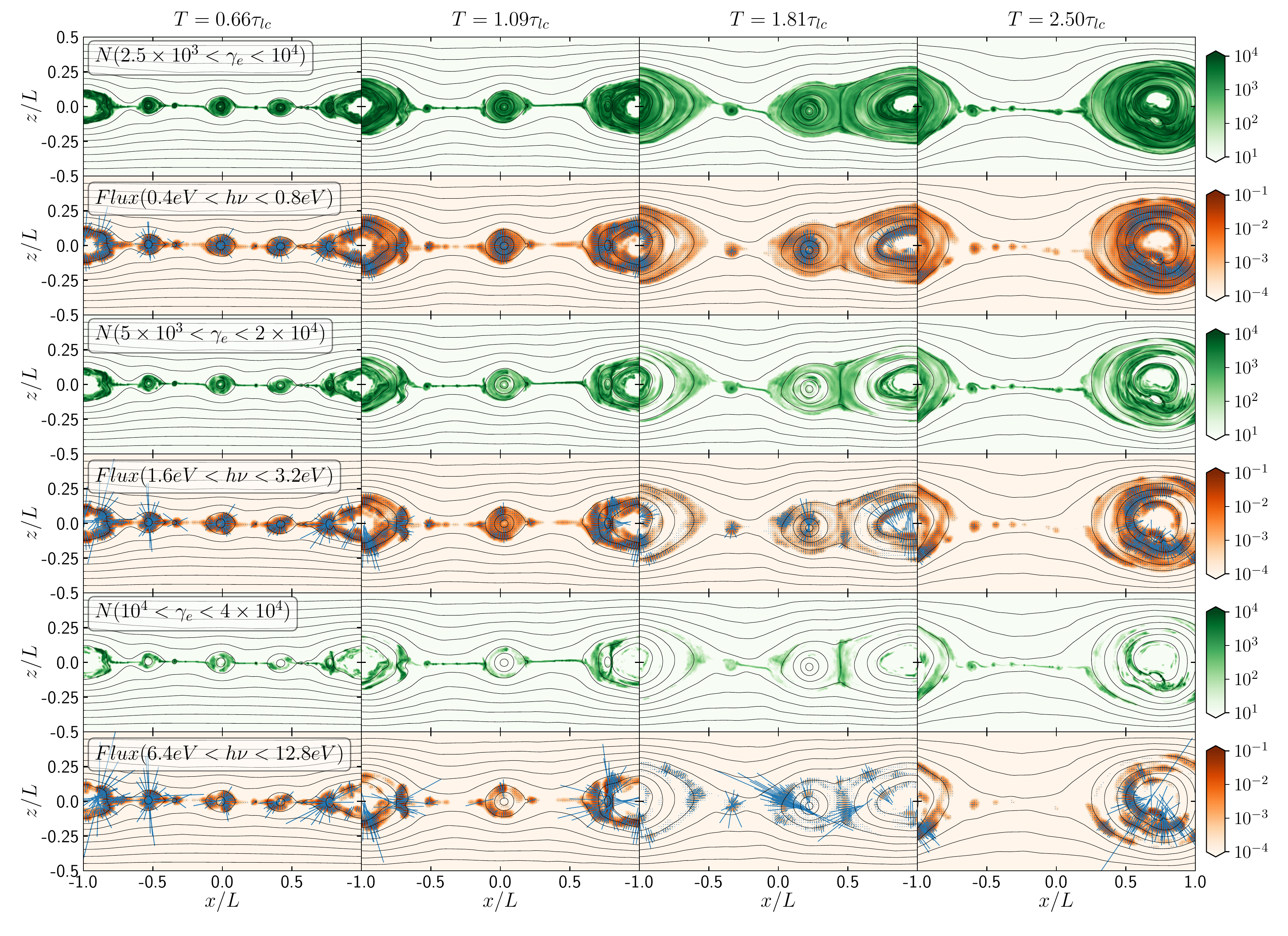}
    \caption{Snapshots of particle spatial distributions of different Lorentz factors and polarized emission maps at corresponding observational bands. The magnetic field snapshots are not plotted, as they are the same for different observational bands.}
    \label{fig:obsband4270116160}
\end{figure*}

In light of the harder-when-brighter trend, we expect that the radiation and polarization signatures can depend on observational bands. Figure \ref{fig:obsband} plots light curves and temporal PD and PA variations for infrared, optical, and ultraviolet bands. They generally represent three positions in the spectrum (refer to Figure \ref{fig:defaultspec}), namely, around cooling break (infrared), in the cooling spectrum (optical), near the spectral cutoff (ultraviolet). We find several interesting patterns in light curves. One is that the higher-energy bands tend to flare earlier than the lower-energy bands. This is because the number of higher-energy particles tops at the maximal efficiency of magnetic reconnection during plasmoid mergers. Since the magnetic field lines are nearly anti-parallel at the contact region of the two merging plasmoids, the acceleration is most efficient and can accelerate highest energy particles. As the merging moves beyond the contact region, it can no longer accelerate the highest energy particles. Additionally, the synchrotron losses cool down these highest energy particles. Since the cooling is proportional to the electron energy, at lower energies, the cooling is slower, so that electrons can continue to accumulate even if the reconnection efficiency drops, resulting in a delay in the flare peak. This can be seen clearly in Figure \ref{fig:obsband4270116160}, where highest energy particles that are responsible for the UV emission mostly exist near the X point and merging contact region, while lower energy particles cover a larger spatial region in the reconnection layer. Another feature is that the higher-energy bands show more spikes in the light curve than lower-energy bands. Additionally, we find that the flare amplitude, which we define as the ratio of flare peak over the flux after the saturation of reconnection (saturation is at $\sim 3\tau_{lc}$ in Figure~\ref{fig:obsband}), is much larger for higher-energy bands. Apparently, both patterns can be attributed to the fact that high-energy particles are short-lived compared to low-energy ones. As we can see in Figure \ref{fig:obsband4270116160}, high-energy particles are present only at the formation of relatively large plasmoids or during major plasmoid mergers. At other times, they cool very fast and have very limited number. However, at their respective peaks, the maximal flux of higher-energy bands is not significantly lower than that of the lower-energy bands, thus the higher-energy bands exhibit larger flare amplitude and more spikes than lower-energy bands.

The average PDs between different observational bands appear very similar. However, higher-energy bands are more variable than lower-energy counterparts. This is more clearly illustrated in the time-dependent PA evolution, where one can see much stronger variations, in particular, large PA swings in ultraviolet. These features are directly related to the spatial distribution of nonthermal particles. Owing to the fast synchrotron cooling, high-energy particles are mostly in very localized regions near the plasmoid merging sites (Figure \ref{fig:obsband4270116160} fifth row), and can only survive a short period of time. As a result, their emission represents the highly dynamical evolution of plasmoid merger events, leading to strongly variable PD and PA. On the other hand, low-energy particles can survive a longer period of time, so that they are distributed in much larger regions in the neighborhood of the reconnection layer (Figure \ref{fig:obsband4270116160} first row). Since the reconnection region has relatively disordered magnetic field structure, the emission by these uncooled particles can contaminate the polarization signatures from large plasmoid mergers. As a result, even the polarized flux from the large plasmoid mergers is not very dominating compared to other parts of the reconnection region (Figure \ref{fig:obsband4270116160} second row, where the relative polarized flux represented by the length of blue dashes is shorter than those in the last row). Therefore, their resulting PD and PA represent the overall evolution of the reconnection layer, which is less variable. Nonetheless, in either case the reconnection plane has very disordered magnetic field morphology, thus the average PDs between different bands are very similar. We notice that after the reconnection saturates at $\sim 3 \tau_{lc}$, the ultraviolet band still shows PA rotations. This is because although reconnection has saturated, occasionally there are still a few small plasmoids are generated, which can merge onto the large plasmoid at the periodic boundary. Due to the strong cooling for these high-energy particles, the emission from the reconnection layer is completely dominated by these small flashes of small plasmoid mergers. This explains why we do not see these signatures at infrared and optical bands, whose emission is dominated by the uncooled particles that contaminate the polarization signatures. Nonetheless, after the reconnection saturates, the flux in all bands is very low, so that these polarization signatures should not be observed.

\subsection{Effects of the Guide Field Strength}

\begin{figure}
    \centering
    \includegraphics[width=\linewidth]{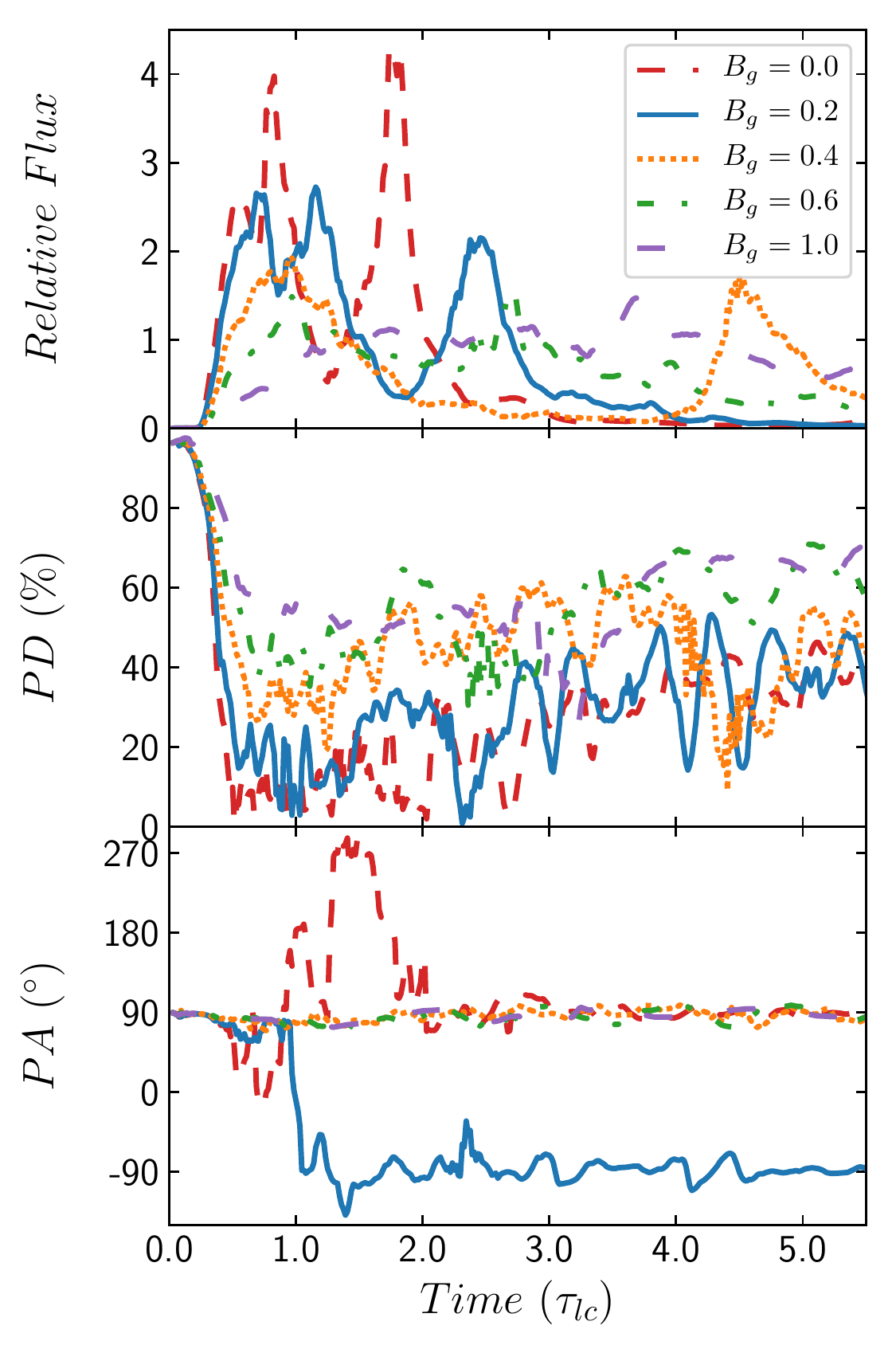}
    \caption{From top to bottom are the optical light curves, temporal PD and PA for different guide fields (they are GF1, default run, GF2, GF3, and GF4, $B_0$ is dropped in the legend).}
    \label{fig:guidefield}
\end{figure}

As shown so far, the time-dependent radiation patterns, especially the polarization signatures, are strongly dependent on the plasmoid motion and mergers. The strength of the guide field plays an essential role on the plasmoid production and evolution \citep[e.g.][]{Ball19,Liu19}. Here we investigate how radiation and polarization signatures can depend on the guide field. Figure \ref{fig:guidefield} shows the results. We can immediately observe several trends here. First, the flare duration, including both individual variability time scale and the overall duration, is shorter for smaller guide fields. Second, the flare amplitude is higher for smaller guide fields \citep{Guo2020}. In the most extreme case where $B_g/B_0=1.0$, the flare amplitude is less than a double, clearly showing that reconnection acceleration is not efficient with large guide fields. Most importantly, we see that the average PD is higher for larger guide fields. In particular, the average PD for $B_g/B_0\gtrsim 0.5$ is very high, at $\gtrsim 40\%$ throughout the reconnection evolution. Given their low flare amplitudes and very high PD, we suggest that magnetic reconnection with large guide field is unlikely responsible for typical blazar variability. Interestingly, for any guide field strengths, the PD is always variable. Lastly, we observe that the PA swings are only present when the guide field is small. Specifically, the $B_g=0$ case exhibits very strong PA rotations, which is consistent with our previous work \citep{Zhang18}.

\begin{figure*}
    \centering
    \includegraphics[width=\textwidth]{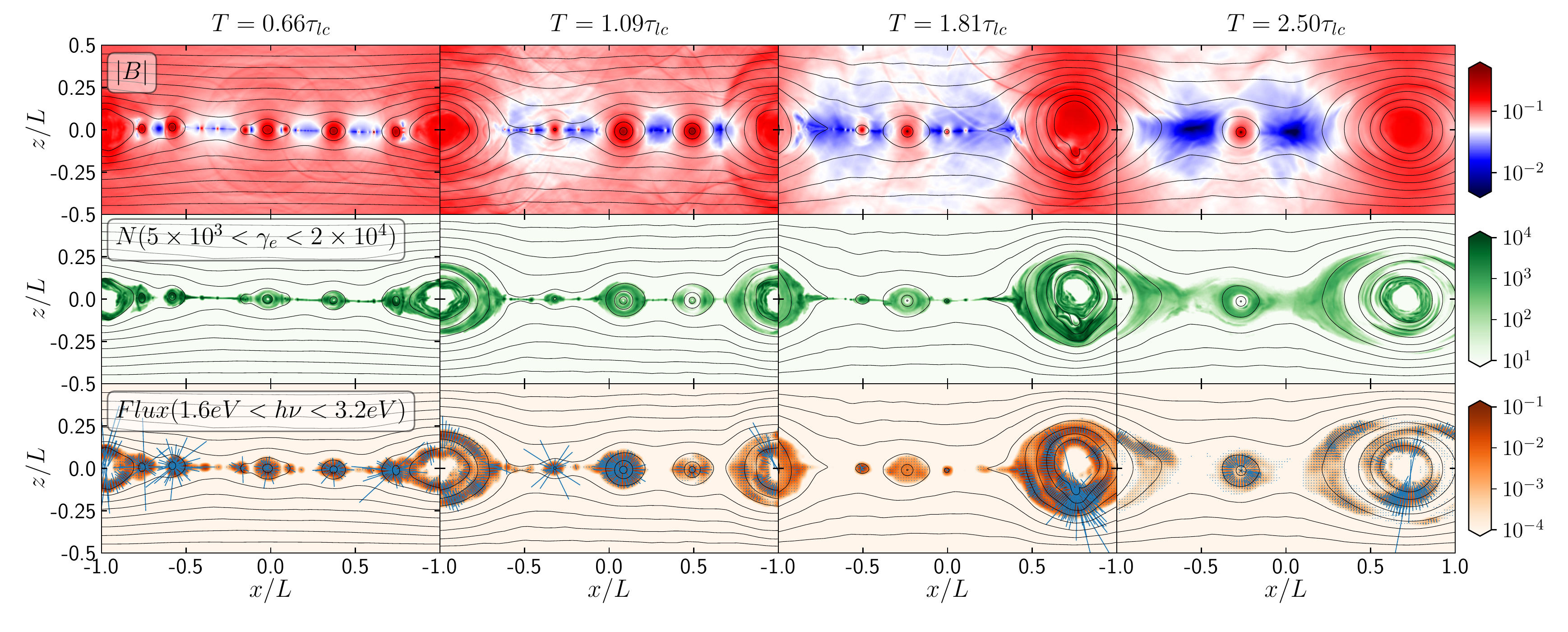}
    \caption{Snapshots of the spatial distributions of magnetic field strengths, nonthermal particles, and polarized emission for the GF1 run ($B_g=0.0$). Blue dashes in the third row represent the relative flux and PA of the polarized emission in each cell.}
    \label{fig:bg004270116160}
\end{figure*}

\begin{figure*}
    \centering
    \includegraphics[width=\textwidth]{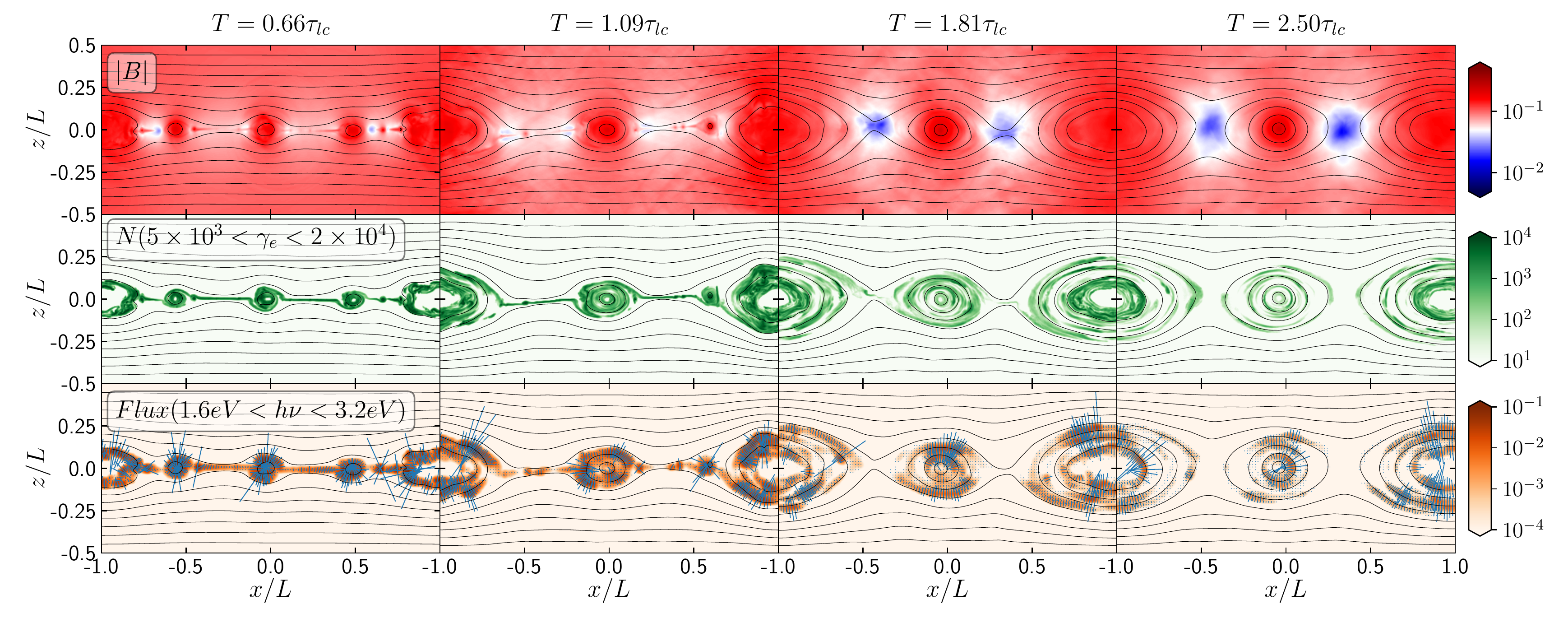}
    \caption{Same as Figure \ref{fig:bg004270116160} but for the GF2 run ($B_g/B_0=0.4$).}
    \label{fig:bg044270116160}
\end{figure*}

\begin{figure*}
    \centering
    \includegraphics[width=\textwidth]{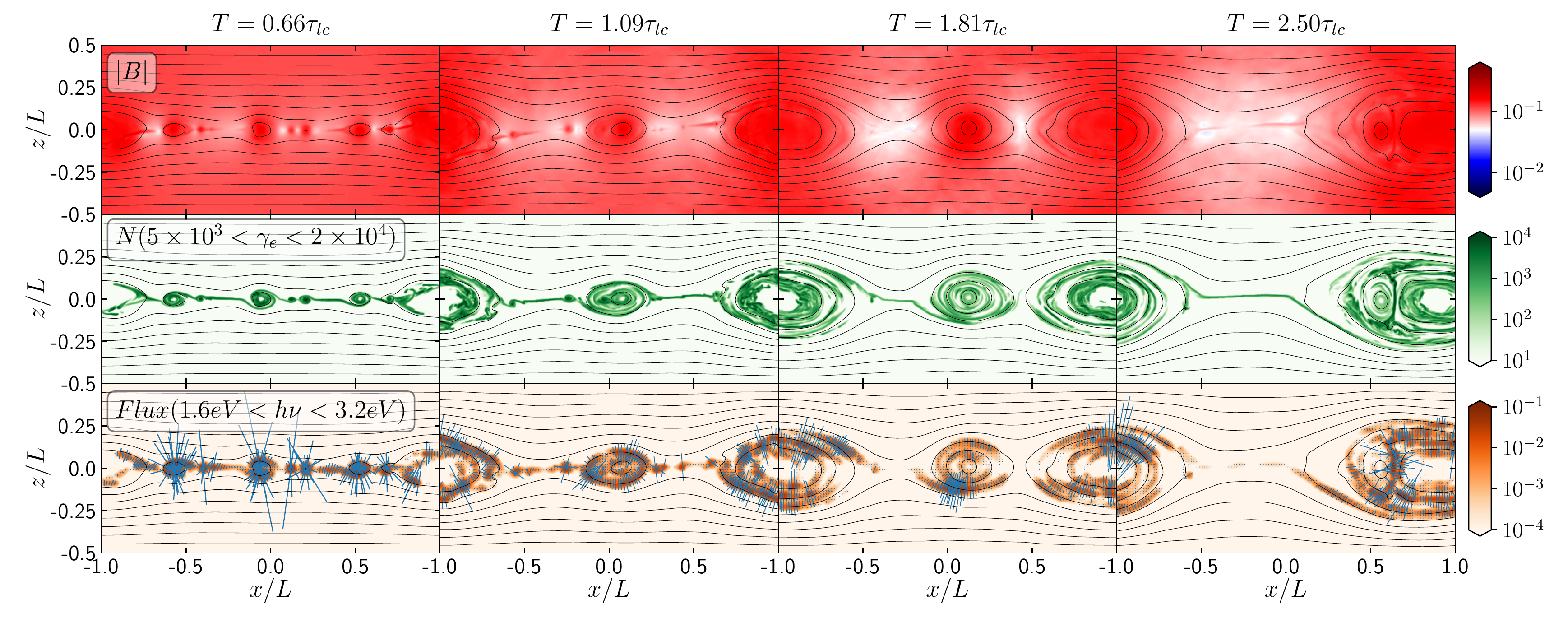}
    \caption{Same as Figure \ref{fig:bg004270116160} but for the GF3 run ($B_g/B_0=0.6$).}
    \label{fig:bg064270116160}
\end{figure*}

\begin{figure*}
    \centering
    \includegraphics[width=\textwidth]{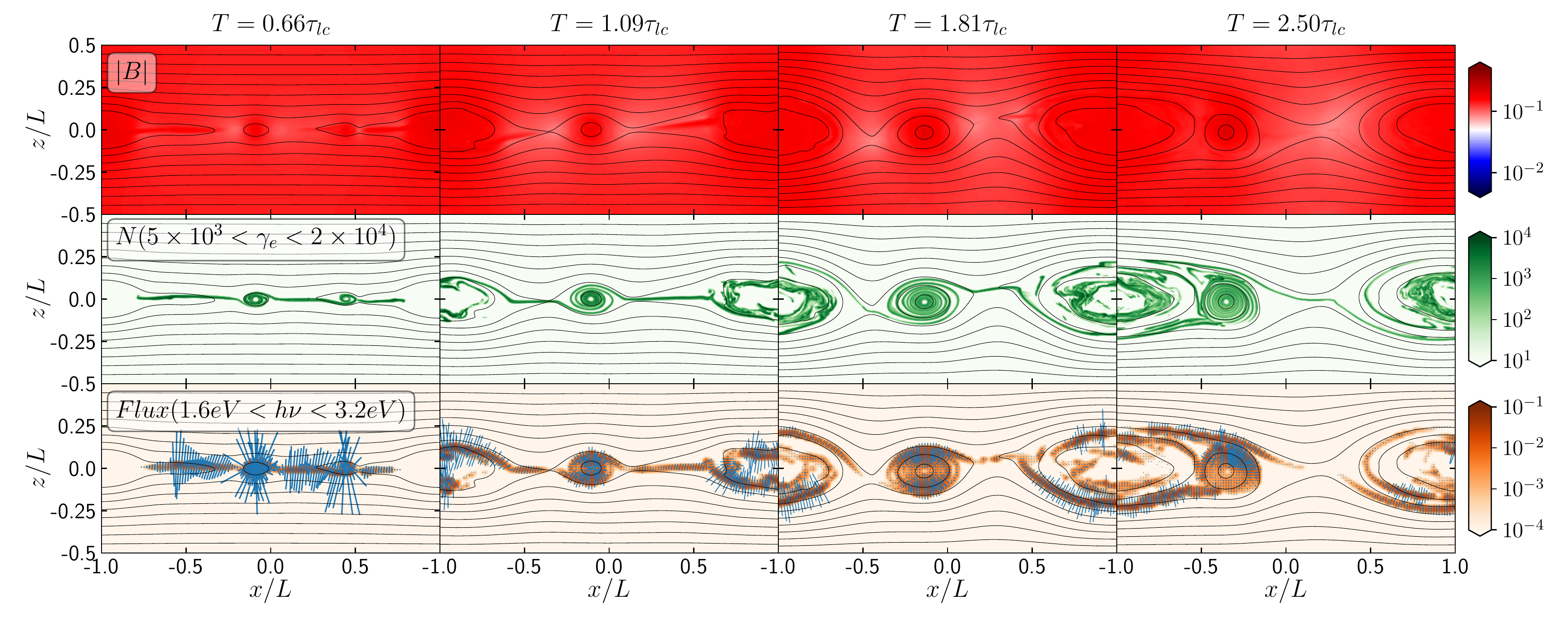}
    \caption{Same as Figure \ref{fig:bg004270116160} but for the GF4 run ($B_g/B_0=1.0$).}
    \label{fig:bg104270116160}
\end{figure*}

The above observational trends originate from the difference in plasmoid production and mergers for different guide field strengths. Comparing the default case (Figures \ref{fig:default3070160256}, \ref{fig:default54586266}, and \ref{fig:obsband4270116160}) with different guide field simulations (Figures \ref{fig:bg004270116160}-\ref{fig:bg104270116160}), we can clearly see that the reconnection development is slower for larger guide fields. This explains the longer flare duration for larger guide field cases. Additionally, more magnetic energy is dissipated to accelerate nonthermal particles in the smaller guide field cases, since there are more nonthermal particles for smaller guide fields (middle row in Figures \ref{fig:bg004270116160}-\ref{fig:bg104270116160}). Apparently, how fast the magnetic energy is dissipated depends on the guide field strength. Consequently, we observe higher flare amplitudes for smaller guide fields \citep{Guo2020}.

Most importantly, the numbers of plasmoids and plasmoid mergers, respectively, are anti-correlated to the guide field strength. This is because a finite guide field can slow down the formation of secondary magnetic islands by preventing the reconnection exhausts from collapsing~\citep{Liu19}. As clearly illustrated in Figures \ref{fig:bg004270116160}-\ref{fig:bg104270116160}, the number of plasmoids at the same time of the simulations is larger with smaller guide fields. Specifically, for $T=1.09\tau_{lc}$ and $T=1.81\tau_{lc}$, we can see that the $B_g/B_0=0.4-1.0$ cases (GF2-GF4) have very little or no new plasmoid production, but for smaller guide fields ($B_g/B_0=0-0.2$, GF1 and default run) still exhibit some plasmoid production and/or merger events. Furthermore, we find that the plasmoid mergers are also weakened by the strong guide field, which reduces the nonthermal particle acceleration at the secondary reconnection site and slows down the overall merger process. We can easily see in Figures \ref{fig:default54586266} and \ref{fig:bg004270116160} third column, where large plasmoid mergers are ongoing, the polarized flux from the merger site is very dominating in both cases (refer to the lengths of the blue polarized flux dashes in the third rows). On the contrary, the polarized flux from plasmoid mergers is not dominating for larger guide field strengths. This is clearly shown in the last column of Figures \ref{fig:bg064270116160} and \ref{fig:bg104270116160}. Consequently, even when large plasmoid mergers happen in a reconnection layer with considerable guide field components, they cannot lead to large PA rotations. Finally, plasmoid production and their mergers make the overall magnetic morphology very disordered, resulting in lower PD with smaller guide fields. Nonetheless, considering that the reconnection is a violent process that significantly alters the magnetic field structure, the PD is always variable for any guide field strengths, which represents the plasmoid evolution in the reconnection plane.

\section{Additional Parameter Studies \label{sec:parameterstudy}}

In the previous section, we have shown that the magnetic reconnection exhibits profound radiation and polarization signatures at different observational bands, which are due to the location of the specific band in the synchrotron-cooled spectrum. Additionally, we find that guide field strengths can strongly affect the reconnection dynamics and the resulting emission. In this section, we perform additional parameter studies to understand how radiation and polarization signatures can depend on other physical parameters in the reconnection region. These parameters are the magnetization factor ($\sigma$), the cooling factor, and the upstream electron temperature. The magnetization factor is clearly a very important parameter that affects the reconnection dynamics, as shown by many previous studies \citep{Sironi14,Guo14,Petropoulou16}. The cooling factor is a free parameter in our simulation that affects the radiative cooling of particles, which may result in observable signatures. The upstream electron temperature may affect the magnetic reconnection dynamics \citep{Petropoulou19}, but, in general, the thermal plasma in the reconnection upstream is not expected to make a considerable contribution to the blazar emission, as shown in most blazar spectral fitting studies \citep{Boettcher13}. Finally, we also study the emission from a larger simulation box, so as to examine if the results are applicable to larger physical sizes, which are necessary in realistic blazar models.

\subsection{Magnetization Factor}

\begin{figure}
    \centering
    \includegraphics[width=\linewidth]{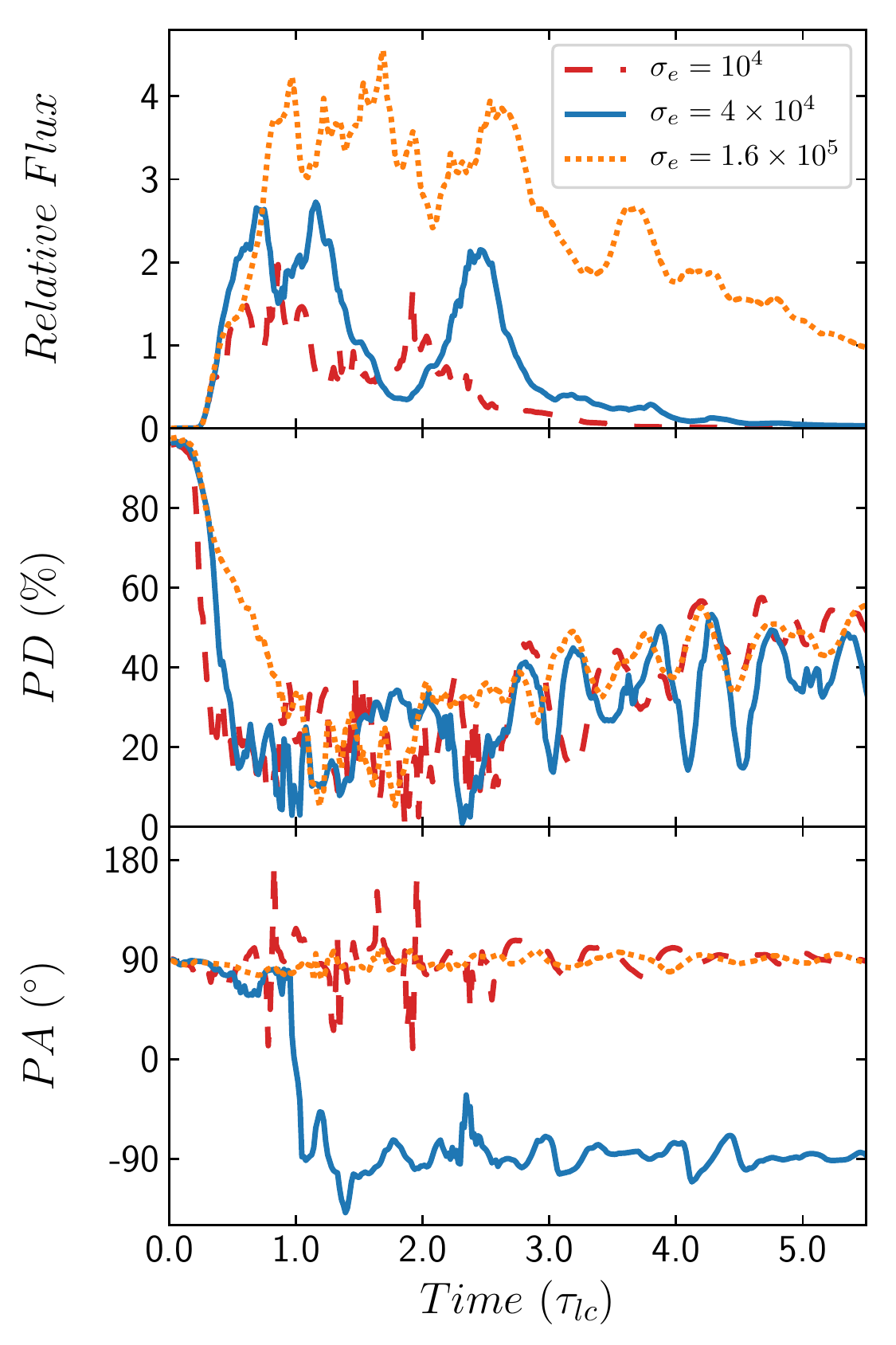}
    \caption{From top to bottom are the optical light curves, temporal PD and PA for different magnetization factors (they are MF1, default run, and MF2). Notice that we rescale the light curves of MF1 and MF2 runs by 5 and 0.5, respectively, to make them appear in the same scale as the default run.}
    \label{fig:sigma}
\end{figure}

\begin{figure*}
    \centering
    \includegraphics[width=\textwidth]{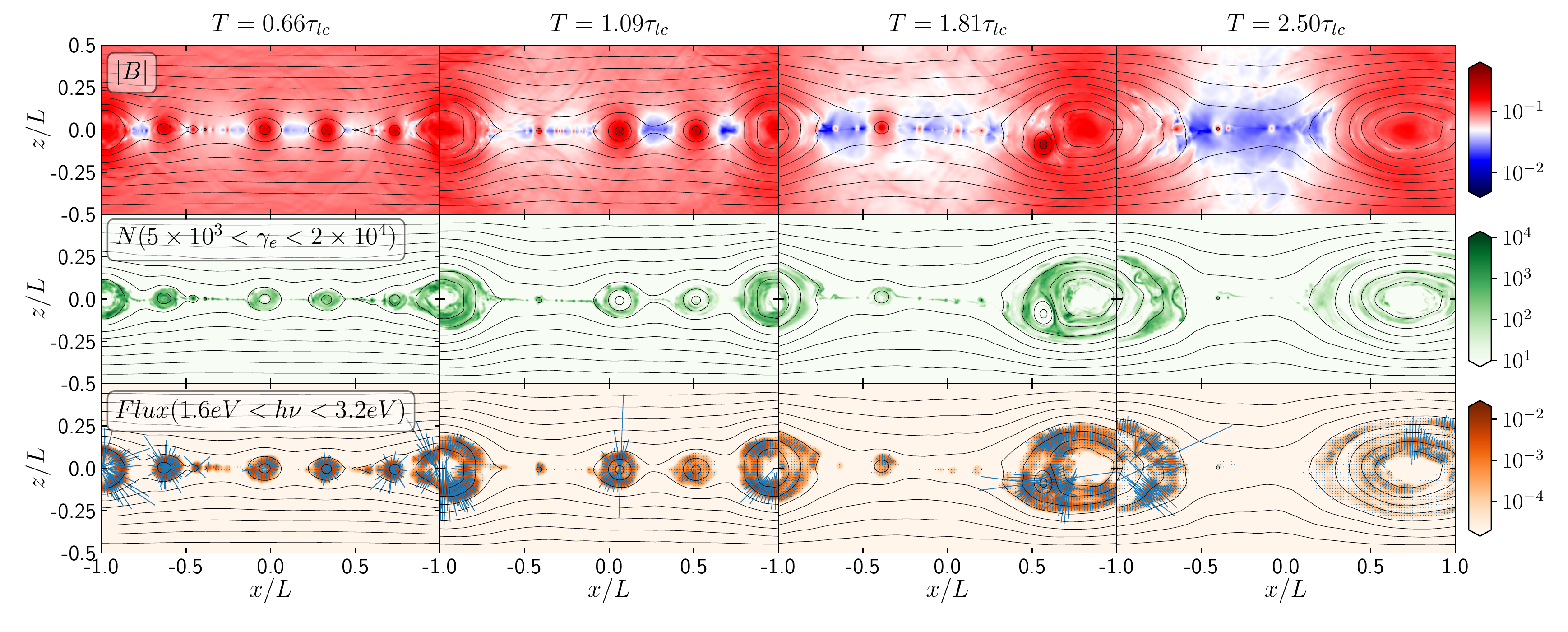}
    \caption{Same as Figure \ref{fig:bg004270116160} but for the MF1 run ($\sigma_e=10^4$).}
    \label{fig:sigma1e44270116160}
\end{figure*}

\begin{figure*}
    \centering
    \includegraphics[width=\textwidth]{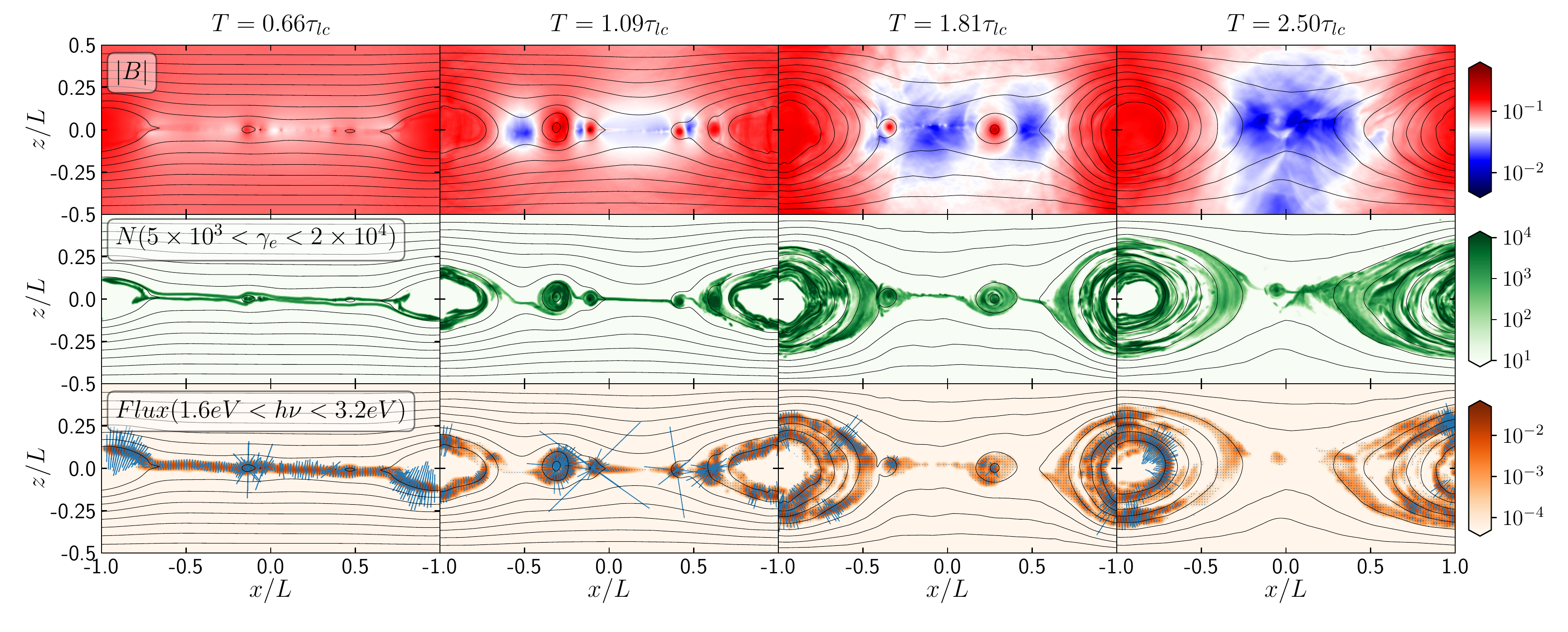}
    \caption{Same as Figure \ref{fig:bg004270116160} but for the MF2 run ($\sigma=1.6\times10^5$).}
    \label{fig:sigma16e44270116160}
\end{figure*}

We consider two other magnetization factors, $\sigma_e=10^4$ and $\sigma_e=1.6\times 10^5$. Since we use proton-electron plasma with real mass ratio, they correspond to $\sigma\sim 5$ and $\sigma\sim 87$, respectively. We normalize the strength of their anti-parallel magnetic field components to the same value, $B_0\sim 0.1~\rm{G}$ at the beginning of the simulation, while the other parameters are kept to their reference values. Comparing these two cases, we find that there are more magnetic energy dissipated with higher magnetization factor (Figures \ref{fig:sigma1e44270116160} and \ref{fig:sigma16e44270116160}). Also we see more nonthermal particles accelerated in the MF2 case. Nevertheless, the overall plasmoid production and merger appear qualitatively similar between the two cases. Comparing to the study of different guide fields in the previous section, apparently the magnetization factor does not play a so important role on the plasmoid evolution as the guide field strength within the parameter regime that we are interested in here.

Figure \ref{fig:sigma} shows the light curves and polarization variations for these cases. Since higher magnetization leads to more particle acceleration, the flux level is strongly affected by the magnetization factor. We also notice that the flare peaks later in the high magnetization run than the low magnetization run. The PDs between the three runs are very similar, but the low magnetization case shows strong PA rotations. These behaviors are very similar to those in the default run at different observational bands. Indeed, this is because the magnetization factor can affect the spectral shape. It has been shown in previous works that the electron power-law cutoff is approximately at $\sigma_e$ for magnetic reconnection \citep[e.g.,][]{Sironi14,Guo14}. Take the $\sigma_e=10^4$ case as an example, the nonthermal electrons cut off at lower energies than the default case, so that the optical band is around the spectral cutoff, similar to the ultraviolet band in the default case. Therefore, we observe highly variable PD and PA in the optical band for the $\sigma_e=10^4$ case, similar to the ultraviolet band in the default run (orange curves in Figure~\ref{fig:obsband}). Clearly, temporal variations alone cannot diagnose the magnetization factor in the reconnection layer, but we need spectral information as well.

\subsection{Cooling Factor}

\begin{figure}
    \centering
    \includegraphics[width=\linewidth]{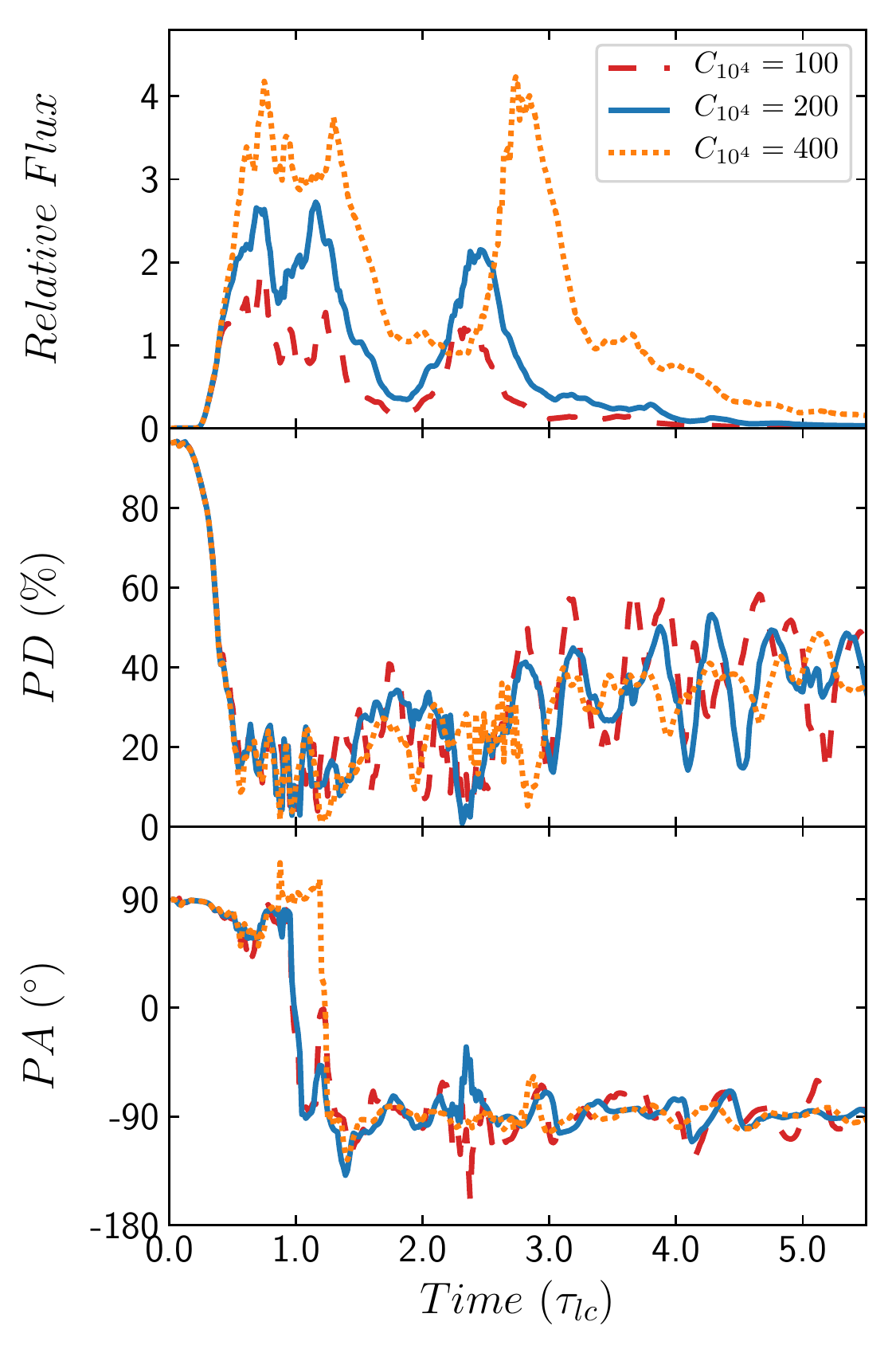}
    \caption{From top to bottom are the optical light curves, temporal PD and PA for different cooling factors (they are CF1, default run, and CF2).}
    \label{fig:cool}
\end{figure}

\begin{figure*}
    \centering
    \includegraphics[width=\textwidth]{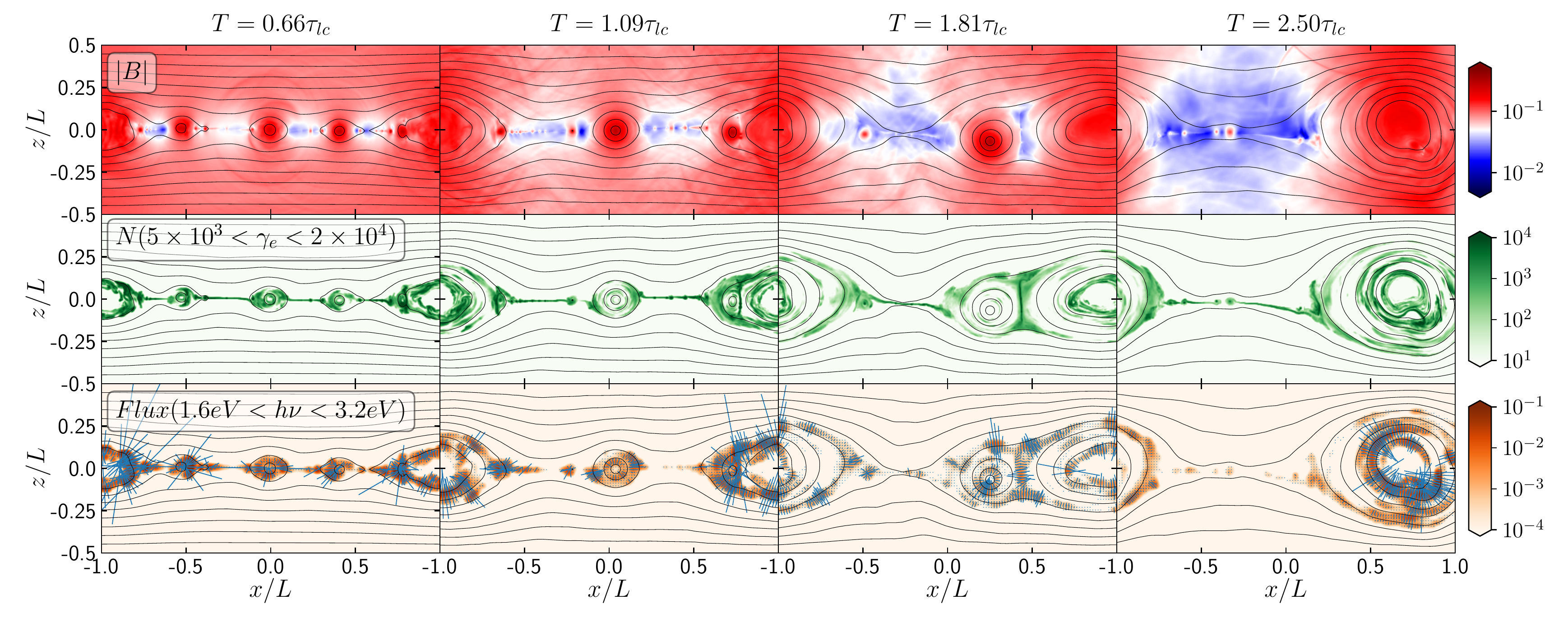}
    \caption{Same as Figure \ref{fig:bg004270116160} but for the CF1 run.}
    \label{fig:cool1004270116160}
\end{figure*}

\begin{figure*}
    \centering
    \includegraphics[width=\textwidth]{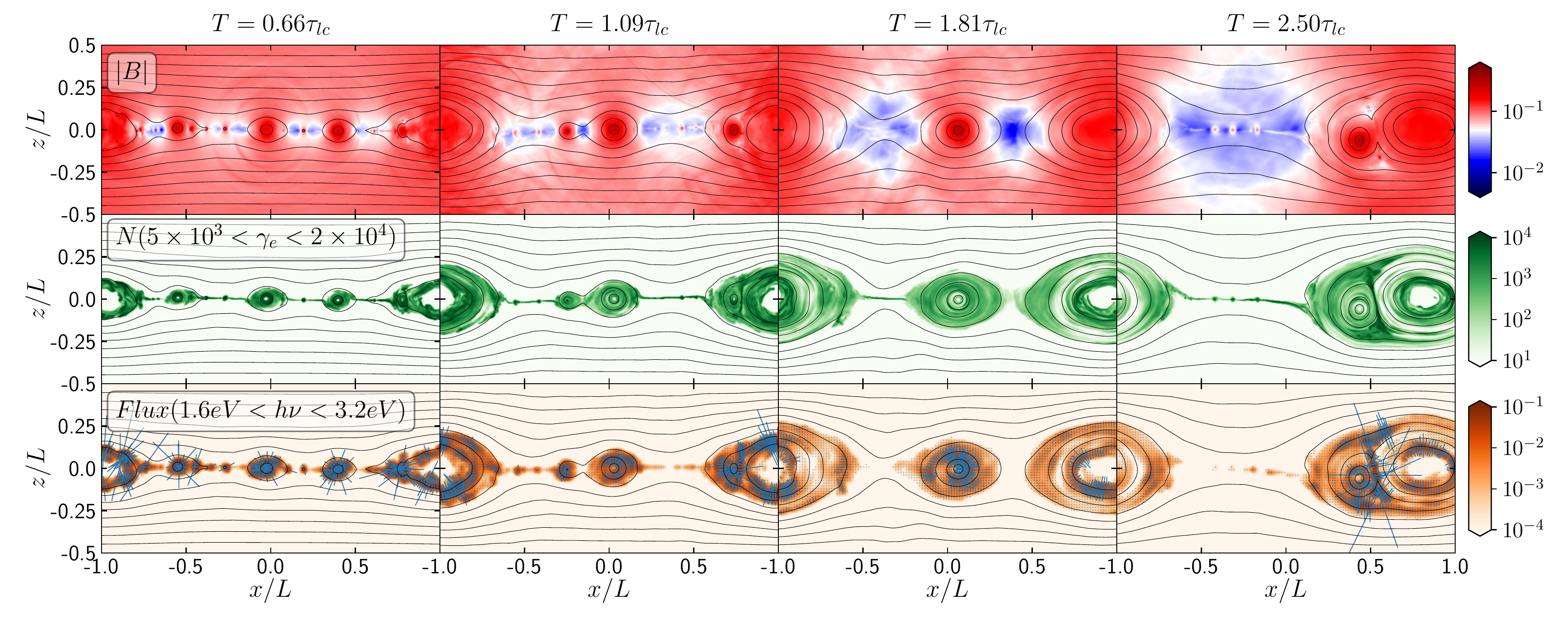}
    \caption{Same as Figure \ref{fig:bg004270116160} but for the CF2 run.}
    \label{fig:cool4004270116160}
\end{figure*}

The cooling factor $C_{10^4}$ describes how fast the radiative cooling time scale is compared to the acceleration time scale. Since it is a free parameter that we manually add into the simulation, it is crucial to examine how this parameter can affect the robustness of our results. Figure \ref{fig:cool} shows the results. We find that the overall flux level is considerably affected by different cooling factors. This is straightforward to understand, as stronger cooling will reduce the number of nonthermal particles in the reconnection region, leading to lower flux. We also find that the reconnection dynamics and plasmoid evolution are remarkably similar between different cooling factors (Figures \ref{fig:cool1004270116160} and \ref{fig:cool4004270116160}), except that the overall evolution is slower for slower cooling. We suggest that the radiative reaction force may play a role in the reconnection dynamics. More importantly, stronger cooling makes the nonthermal particles cool faster, very similar to the different cooling experienced by different observational bands in the default run. The above two effects lead to a delay in the light curves in Figure \ref{fig:cool}. We observe that the average PD is very similar for different cooling factors, consistent to the different observational bands in the default run. This means that the overall magnetic field morphology and evolution is not affected by the radiative reaction force beyond the evolution rate. Furthermore, we do not observe major difference in the PA evolution. The reason turns out to be that the ratio between the three cooling factors is relatively small, so that the optical band is at similar location in the spectrum as the default case, i.e., between the cooling break and the spectral cutoff. We also find that the number of nonthermal particles is fewer for smaller cooling factor (the CF1 run, which means stronger radiative cooling), but the difference is not so large compared to the nonthermal particles that are responsible for different observational bands in the default case (Figure \ref{fig:obsband4270116160}). This is understandable, as the cooling factor is only varied by a factor of 2 in our studies. Nonetheless, we have examined that for $2\times$ faster cooling ($C_{10^4}=50$), the optical PD and PA evolution become very different from the default case, qualitatively similar to the ultraviolet curve in the default run in Figure \ref{fig:obsband}. Based on the discussion in this section and the previous section on the different observational bands, we conclude that the polarization variation patterns strongly depend on the position of the observational band on the blazar SED. Specifically, if the observational band is close to the spectral cutoff, the PD and PA can be highly variable; but if it is around or before the cooling break in the SED, the PD and PA generally show small and erratic fluctuations around some average value.

\subsection{Upstream Temperature}

\begin{figure}
    \centering
    \includegraphics[width=\linewidth]{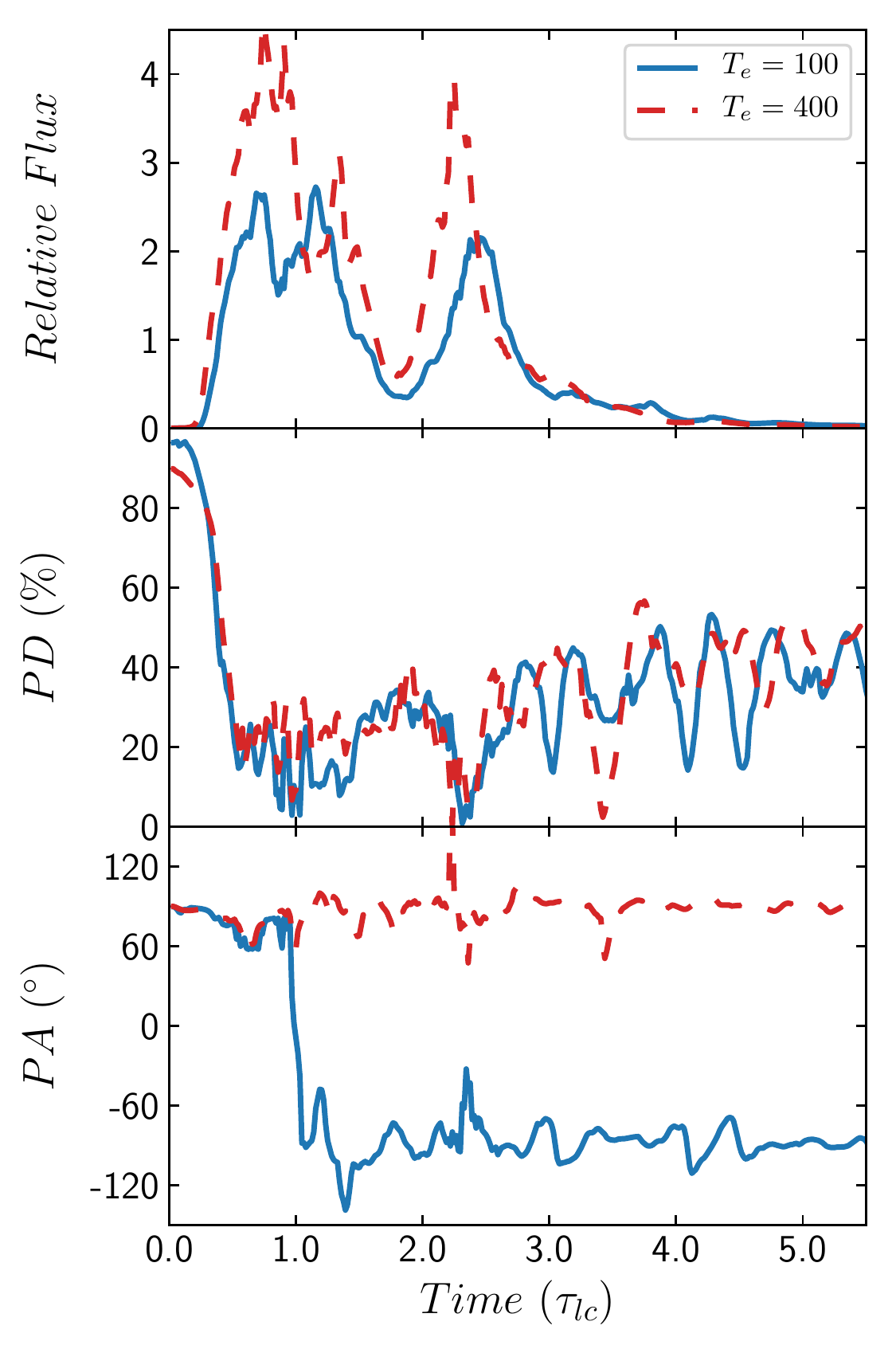}
    \caption{From top to bottom are the optical light curves, temporal PD and PA for different upstream temperatures (they are default run and UT1).}
    \label{fig:vthe}
\end{figure}

\begin{figure*}
    \centering
    \includegraphics[width=\textwidth]{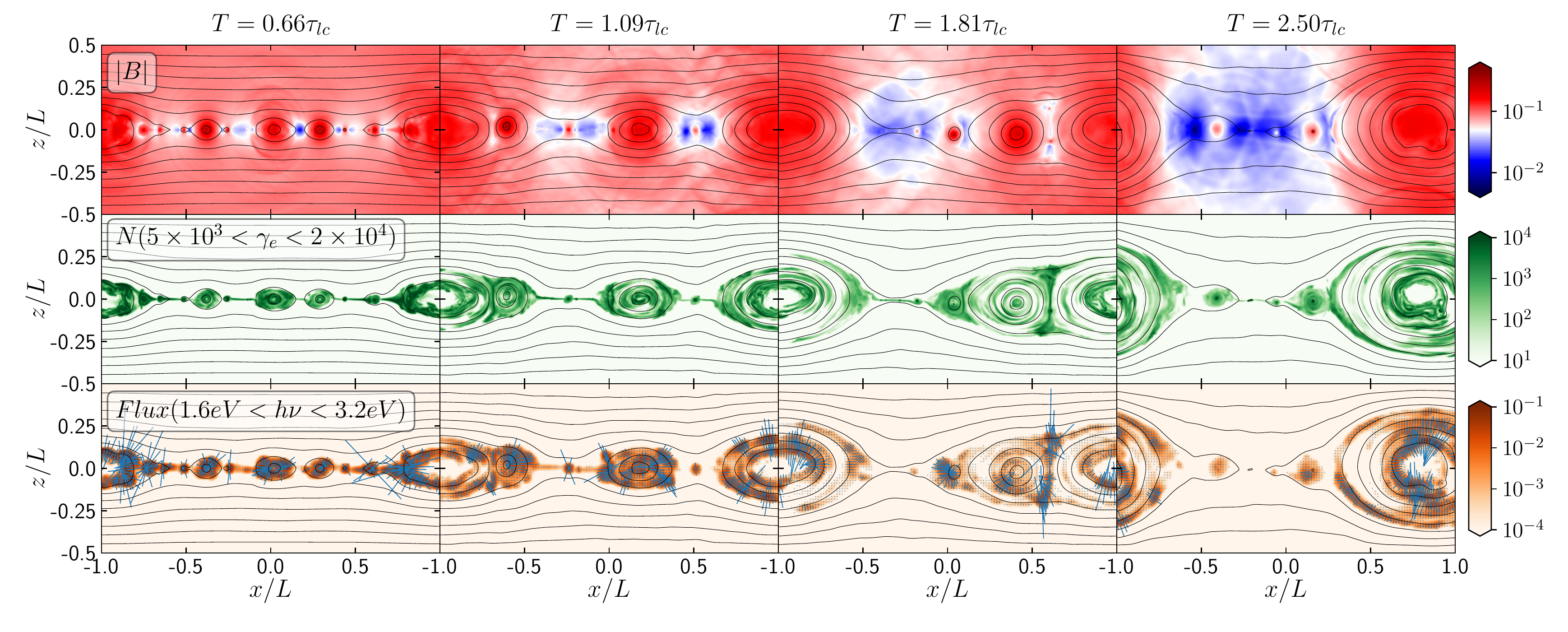}
    \caption{Same as Figure \ref{fig:default3070160256} but for the UT1 run.}
    \label{fig:vthe4004270116160}
\end{figure*}

Since the blazar zone is a highly energetic region, electrons therein may be significantly heated. As suggested by blazar spectral fitting models, the upstream electron temperature, which is often modeled as the low-energy electron spectral cutoff, can be as high as a hundred to a few thousand $m_ec^2$ \citep[e.g.,][]{Boettcher13,Paliya18}. Additionally, since the PIC simulation needs to resolve the thermal electron inertial length $d_e$, smaller upstream temperature requires higher resolution. Thus for practical reasons we just compare two different upstream electron temperate, $T_e=100$ (the default case) and $T_e=400$ (Figure \ref{fig:vthe}). Apparently, the light curves and PD are very similar between the two cases, except that the $T_e=400$ case shows higher flux. This is likely due to that the thermal particles start from higher energies, so that more nonthermal particles are accelerated. In the PA evolution, the $T_e=400$ case does not show a $\gtrsim 180^{\circ}$ swing at the second flare, but it does show a PA rotation of $> 90^{\circ}$ during its last flare. This suggests that the large plasmoid mergers can still dominate the emission in the reconnection layer. As shown in Figure \ref{fig:vthe4004270116160} the nonthermal particle evolution and polarized flux appear remarkably similar to the default run (Figure \ref{fig:obsband4270116160} third and fourth rows). The lack of $\gtrsim 180^{\circ}$ PA swing during the second flare is probably because the large plasmoid mergers happen to have symmetric production of nonthermal particles in clockwise and counterclockwise directions. We notice that \cite{Petropoulou19} have found that the different upstream temperatures may affect the reconnection dynamics. But their studies have investigated drastically different upstream temperatures. In our parameter study, we only vary the upstream temperature by a factor of 4. Apparently, it only slightly affects the particle acceleration during large plasmoid mergers at the second flare. Therefore, we suggest that the upstream temperature does not play a crucial role as other parameters in radiation and polarization signatures from reconnection.

\subsection{Box Size}

\begin{figure}
    \centering
    \includegraphics[width=\linewidth]{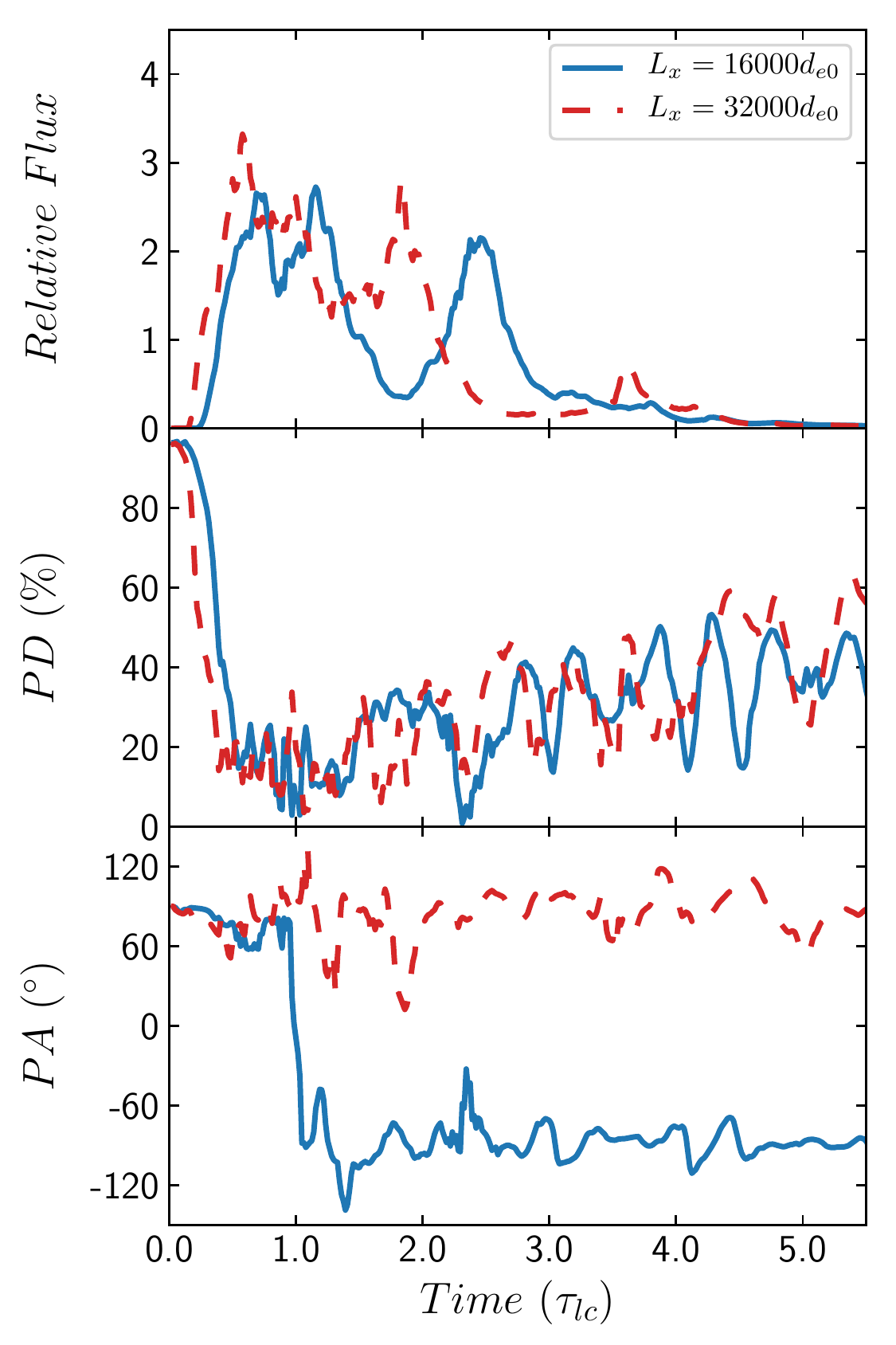}
    \caption{From top to bottom are the optical light curves, temporal PD and PA for different simulation box sizes (they are default run and BS1). The light curve of the larger box case is rescaled by a factor of 0.5 to be in the same figure.}
    \label{fig:box}
\end{figure}

\begin{figure*}
    \centering
    \includegraphics[width=\textwidth]{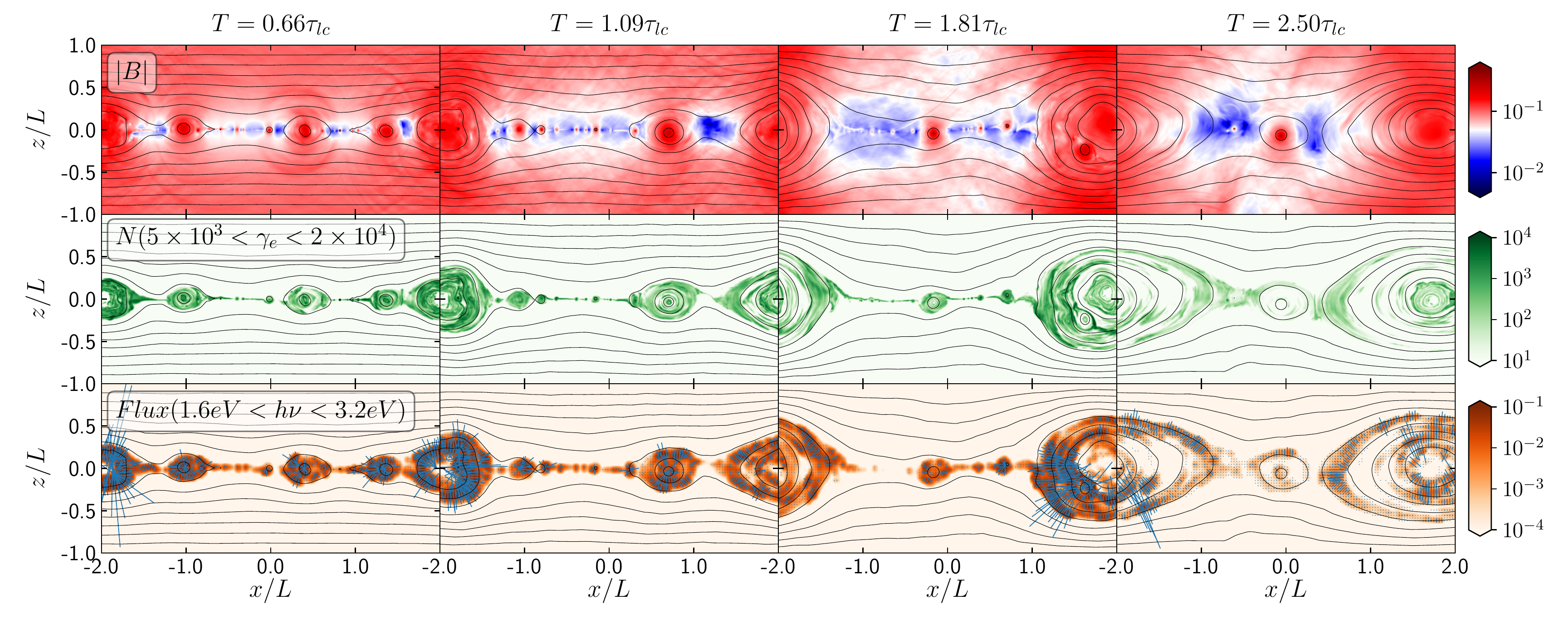}
    \caption{Same as Figure \ref{fig:default3070160256} but for the BS1 run.}
    \label{fig:large4270116160}
\end{figure*}

The physical processes that are responsible for the time-dependent radiation signatures, especially the polarization signatures, are the production of plasmoids and secondary reconnection at the plasmoid mergers. As we can see in the simulations, these processes are not subject to the particle kinetic scales that PIC simulations mostly deal with. Nevertheless, it is very important to examine if these patterns may change with a larger simulation box. We pick a box that is twice as big as the default case in both $x$ and $z$ directions, and redo the simulation with the same parameters as the default run. Figure \ref{fig:box} shows the results. Apparently, all observable signatures, including polarization variations, are very similar to the default case. We remind the readers that for both time-dependent signatures and snapshots of the simulation domain we use the light crossing time scale as the time unit. Now that the simulation box is twice as large, the light crossing time scale is also twice as large. Therefore, the snapshots in Figure \ref{fig:large4270116160} are taken at twice the evolution time as the default case, which represent more mature states in the reconnection evolution. This is also evident by the light curve and polarization variation, where the signatures appear very similar to the default case if we stretch them by a factor of two. The readers may notice that the large box simulation does not show a $\gtrsim 180^{\circ}$ PA swing as the default case. We remind the readers that large PA swings require continuous plasmoid mergers that have asymmetric nonthermal particle acceleration at the secondary reconnection site in the same direction. As mentioned in the default run, the large PA swing is due to two large plasmoid mergers in which both have more nonthermal particle streaming in the clockwise direction. In the large box simulation, however, we find some plasmoid mergers happen to have more particles streaming in the counterclockwise direction. For instance, the readers may refer to the third column in Figure \ref{fig:large4270116160}. There is a large plamoid merger ongoing at the right end of the simulation box, which has more nonthermal particles streaming counterclockwise, where we find considerable polarized flux in the lower half of the simulation domain. Additionally, we can see in Figure \ref{fig:box} that between $\tau_{lc}$ and $2\tau_{lc}$, there are multiple $\sim 90^{\circ}$ PA swings going up and down, indicating that the direction in which the nonthermal particles are streaming during different large plasmoid mergers keeps changing.

\section{Implications for Observations \label{sec:observation}}

Relativistic magnetic reconnection can widely exist in magnetized plasma in astrophysical systems. In particular, current theories suggest that relativistic jets are launched with very high magnetic energy, which may dissipate to accelerate particles and radiate along its propagation. Our systematic numerical studies on observable signatures from magnetic reconnection have shown intriguing patterns, especially in the polarization variations. Thanks to the comprehensive multi-wavelength coverage and detailed optical polarization monitoring programs, blazar observations have collected a large amount of simultaneous multi-wavelength data with optical polarization signatures \citep[e.g.,][]{Marscher10,Blinov15,Blinov18}. Here we discuss several potential observable patterns that can be used to identify and diagnose magnetic reconnection in blazar jets.

We find an overall harder-when-brighter trend from reconnection. In blazar observations, the harder-when-brighter trend is frequently seen in all wavelengths, including both the synchrotron spectral component and the high-energy spectral component \citep{Giommi90,Abdo10,Krauss16}. As shown in Figure \ref{fig:defaultspec}, the particle spectra show this trend as well. Therefore, we expect that in a leptonic model, where the high-energy component is due to the Compton scattering of the same electrons that make the synchrotron emission, we should expect the same harder-when-brighter trend in X-rays and $\gamma$-rays.

Magnetic reconnection exhibits characteristic radiation and polarization trends in multi-wavelength observations. These include earlier flare peaks for higher-energy bands in the synchrotron spectral component, and more variable polarization signatures, in particular, large PA swings. Apparently, the difference of polarization signatures between different observational bands depend on the position of the observational band on the blazar SED. The time delay between peaks in higher-energy bands and lower-energy bands is apparently proportional to the energy difference between the two bands. For flat-spectrum radio quasars (FSRQs), this delay may be hard to detect, given that the typical FSRQ spectrum starts to cut off beyond optical bands. However, for high-frequency-peaked BL Lacs (HBLs), the X-ray may peak considerably earlier than the optical band. Furthermore, future X-ray polarimeters, such as {\it IXPE} that is scheduled to launch in 2021, may detect X-ray polarization signatures for bright BL Lacs such as Mrk~421 and Mrk~501. This study finds highly variable X-ray polarization signatures if reconnection drives the flares in HBLs. However, considering the sensitivity of {\it IXPE} and other proposed X-ray polarimeters, it is very likely that these instruments have to integrate over longer timescales to obtain one polarization data point. During this time, if the flare is driven by magnetic reconnection, the PA may have considerably rotated. By integrating these photons, those with perpendicular PA can cancel out their polarization signatures. On the contrary, previous works suggest that the shock scenario expects higher PD in the X-ray bands because of the more ordered magnetic field structure at the shock front \citep{Tavecchio18}. Therefore, if X-ray polarimeters detect significantly lower PD in X-ray bands than optical bands in most blazar flares, this can be strong evidence of magnetic reconnection in blazars. 

Based on our results, large optical PA swings are uniquely associated with small guide fields ($B_g/B_0\lesssim 0.2$). Previous works have suggested that current sheets may form via magnetic instabilities/turbulence in the jet, or between oppositely oriented stripes in the striped jet \citep{Begelman98,Giannios06,Giannios19}. The major difference between instabilities/turbulence and striped jet lies in two aspects. First, if kink instabilities happen on the scale of the blazar zone environment \citep{Mizuno09,Guan14,BarniolDuran17}, the current sheets formed therein can be of smaller than the size of the blazar zone. Therefore, the radiation and polarization signatures that we present in this paper can happen on very short time scales. Interestingly, such micro-variability patterns have been reported in observations \citep[e.g.,][]{Pasierb19}. On the other hand, in a striped jet morphology, the magnetic stripes are created at the central engine and propagate out along with the jet. Therefore, the blazar zone can be the location of strong dissipation of magnetic energy within the large current sheets formed between stripes \citep{Giannios19}. In this situation, we expect that the variability in radiation and polarization signatures should be of the typical days to a few weeks blazar flare duration \citep{Marscher10,Angelakis16}. The second difference is that kink instabilities or turbulence are unlikely to form nearly perfectly anti-parallel magnetic field lines, due to their very disordered magnetic topology. Therefore, we do not expect any large ($\gtrsim 90^{\circ}$) PA rotations from reconnection in kink instabilities or magnetic turbulence. Very interestingly, the blazar micro-variability, which may originate from reconnection in magnetic instabilities/turbulence, mostly shows small PA fluctuations \citep{Pasierb19}. In contrast, in the striped jet scenario, a large region of anti-parallel magnetic field lines with small guide fields is probable. Therefore, large PA swings in blazar jets may point to a striped jet model.

Additionally, we find that these PA swings are accompanied by strong flares. Furthermore, the PD generally stays at a low level during PA swings, but can reach higher levels outside swings. Very interestingly, these features have been reported by the RoboPol team, where they find that PA swings are always accompanied by {\it Fermi} $\gamma$-ray flares and the PD is about 50\% during PA swings than quiescent state \citep{Blinov16}. Indeed, while PA swings have been reported in many observations, they are rather rare and extreme observational phenomena, which imply very lucky situations such that the magnetic field lines are well anti-parallel in the reconnection event. RoboPol program has classified blazars as ``rotators'' and ``non-rotators'', in which the former has shown PA rotations \citep{Blinov16b}. They have found that rotators appear more active in radiation and polarization signatures than non-rotators. Based on our simulations, this behavior can be attributed to that rotators are likely striped jet, where reconnection between nearly perfectly anti-parallel magnetic field lines is more likely to happen. Consequently, both light curves and polarization signatures appear more variable. We suggest that with better observational statistics, optical polarization signatures can diagnose the physical conditions of magnetic reconnection in jets as well as the overall jet morphology and dynamics.

\section{Summary and Discussion \label{sec:discussion}}

To summarize, we have presented a systematic study of the radiation and polarization signatures arising from magnetic reconnection in an electron-ion plasma in the blazar zone environment. Our studies are based on first principles via PIC simulations, and cover all spectral and temporal radiation and polarization signatures through polarized radiation transfer simulations. Our studies demonstrate that the most crucial physical processes during reconnection that affect radiation and polarization signatures are the plasmoid production and mergers. In fact, these processes are also unique to magnetic reconnection compared to other blazar flare models including shocks, kink instabilities and magnetic turbulence. In particular, we have shown that PA rotations are linked to large plasmoid mergers in the reconnection. Our systematic studies have explored radiation and polarization signatures for different observational bands,
guide field strengths, magnetization factors, radiative cooling, upstream temperature, and simulation box size. Based on our results, we have discovered several key observable features in the synchrotron spectral component of blazars from magnetic reconnection:
\begin{itemize}
    \item There is a harder-when-brighter trend in the spectral evolution.
    \item Observational bands of lower frequencies (infrared to optical) tend to peak later than those of higher frequencies (ultraviolet to X-ray).
    \item Lower frequency bands also show less variable polarization signatures.
    \item PA swings are unique to magnetic reconnection with small guide fields ($B_g/B_0\lesssim 0.2$).
    \item PA swings appear simultaneous with flares.
    \item The PD generally drops during PA swings.
    \item Reconnection with large guide fields ($B_g/B_0\gtrsim 0.5$) yields low flare levels and very high PD ($\gtrsim 40\%$), which are inconsistent with typical blazar observations.
\end{itemize}

We have shown that the magnetization factor and radiative cooling play important roles in the spectral shape. Specifically, the difference in the radiation and polarization signatures between different observational bands depends on the location of these bands on the synchrotron spectral component of the blazar. The most important physical parameter on the temporal radiation and polarization signatures is the guide field strength. It is particularly important for polarization signatures. Finally, we have shown that the general radiation and polarization patterns are kept for different simulation box sizes, indicating that our findings may be ``scaled up'' to realistic physical sizes of the blazar zone environment.

Generally, current sheets can form in magnetic instabilities/turbulence in the jet, or in a striped jet morphology. We expect that the former may result in short time scale variability in radiation and polarization signatures, because of the highly disordered and dynamical magnetic topology in the instabilities/turbulence; while the latter may lead to strong PA swings and appear very active in both radiation and polarization signatures, since it may give rise to more anti-parallel magnetic field lines.

Our studies are based on 2D PIC simulations. Generally speaking, in reality several 3D effects can influence the radiation and polarization signatures. One obvious factor is the viewing angle, which can strongly affect the light crossing time and the projection of the magnetic field lines that are crucial to both radiation and polarization signatures \citep{Hosking2020}. These effects have to be thoroughly studied with 3D PIC simulations. Another issue is that previous 3D PIC simulations have shown that plasmoids in 2D simulations can extend to flux ropes in 3D \citep{Guo14,Guo15}. However, if the guide field is very weak, the 3D flux ropes can quickly fragment into smaller structures, which we expect that they can lead to similar radiation and polarization signatures as plasmoids \citep{Guo2020b}. On the other hand, if the guide field is strong, it can stabilize the flux ropes. Nonetheless, reconnection is inefficient with strong guide field, and we have shown that strong guide fields can result in very high PD, which are inconsistent with observations. Therefore, we expect that the general radiation and polarization patterns should remain the same in 3D if we are viewing along the guide field direction as in our simulations. Finally, the 3D reconnection can lead to stronger turbulence. This may lead to even lower average PD compared to what we find here.

\acknowledgments{We thank the anonymous referee for very helpful and constructive reviews. This research is supported by Fermi Guest Investigator program Cycle 11, grant number 80NSSC18K1723. HZ, DG, and LD acknowledge supports from the NASA ATP grant NNX17AG21G and the NSF AST grant AST-1816136. XL and FG are grateful for support from DOE through the LDRD program at LANL and DoE/ OFES support to LANL, and NASA ATP program through grant NNH17AE68I. XL and Y.-H.L. are grateful for supports from the National Science Foundation grant PHY-1902867 through the NSF/DOE Partnership in Basic Plasma Science and Engineering. Simulations are conducted on LANL's Institutional Computing machines and Purdue Rosen Center for Advanced Computing (RCAC) clusters. Simulation data can be obtained by emailing HZ or DG.}

\bibliography{PIC+Pol2}
\bibliographystyle{aasjournal}



\end{document}